\documentclass[copyright,creativecommons]{eptcs}    
\providecommand{\event}{QAPL 2010}
\usepackage{breakurl}
\usepackage{amsmath,amssymb,amsfonts,amsthm} 
\usepackage{graphicx}                 
\usepackage{color}                    
\usepackage{enumerate,boxit}
\usepackage{pepa}
\usepackage[square,numbers]{natbib}
\usepackage[font={bf,small}]{caption}
\usepackage[font={bf,scriptsize}]{subfig}
\usepackage[rflt]{floatflt}
\usepackage{fancyvrb}
\usepackage[]{url}

\DefineVerbatimEnvironment%
{code}{Verbatim}
{frame=lines,numbersep=3pt,xleftmargin=10mm,xrightmargin=10mm,fontfamily=tt,fontsize=\small}

\usepackage{preamble}

\def\GPEPA{{GPA}}
\allowdisplaybreaks 

\title{A new tool for the performance analysis of massively parallel computer systems}
\author{Anton Stefanek 
\qquad\qquad 
Richard A.\ Hayden 
\qquad\qquad 
Jeremy T.\ Bradley\thanks{The authors are funded by the EPSRC on the AMPS project
 (reference EP/G011737/1).}
\email{\{as1005,rh,jb\}@doc.ic.ac.uk}
\institute{Department of Computing, Imperial College London}
}

\def\titlerunning{A new tool for the performance analysis of massively parallel computer systems}
\def\authorrunning{A.\ Stefanek, R.A.\ Hayden, J.T.\ Bradley}

\begin{document}

\bibliographystyle{ieeetr}

\maketitle

\begin{abstract}
  We present a new tool, \GPEPA, that can generate key performance
  measures for very large systems. Based on solving systems of
  ordinary differential equations (ODEs), this method of performance analysis
  is far more scalable than stochastic simulation.  The \GPEPA\ tool
  is the first to produce higher moment analysis from differential
  equation approximation, which is essential, in many cases, to obtain
  an accurate performance prediction. We identify so-called {\em
    switch points} as the source of error in the ODE approximation. We
  investigate the switch point behaviour in several large models and
  observe that as the scale of the model is increased, in general the
  ODE performance prediction improves in accuracy. In the case of the
  variance measure, we are able to justify theoretically that in the
  limit of model scale, the ODE approximation can be expected to tend
  to the actual variance of the model.
\end{abstract}

\section{Introduction}

Quantitative analysis of systems by means of differential equation
(ODE) approximation~\cite{pepa-odes-ck,Bor07} or fluid
techniques~\cite{Hil05,Car08} produce transient performance measures
for massive state-space process models of $10^{100}$ states and
beyond. Previous explicit state-space performance techniques which
analysed the underlying continuous-time Markov chain directly (for
example,~\cite{Kno99,Kwi02a}) were limited to $10^{11}$ states in the
very best cases and then only for the simplest steady-state style of
analysis.

In physical and biological processes, deterministic approximation of
system evolution via systems of differential equations have existed
for some
time~\cite{gillespieBook,stochasticProcessesPhysicsChemistry,momentClosure}. However,
differential equation-based techniques have only recently been brought
to bear on process models of computer systems. A major difference
between the two lies in the model of interaction assumed in the two
distinct fields. The model of synchronisation used in computer and
communication systems differs from that typically used in physical and
biological processes, where for example {\em mass-action} dynamics
govern the system evolution.\footnote{This also applies to other
  kinetic laws where the rate function is smooth.} This difference
significantly changes the nature of the differential equation
approximation of computer systems, thus results from mass-action-based
systems cannot be translated to systems based on, for example, process
algebras, queueing networks or stochastic Petri nets.

In particular, there is not as yet a detailed understanding of how
good the differential equation approximation is to the underlying
discrete Markov process as generated from process models of computer
systems. For instance~\cite{Hil05} produced a first order
approximation to large scale Markov models, but there is no discussion
of accuracy of the approximation or higher moment generation.
The issue focuses around so-called {\em switch points} in a
model. In the case of PEPA, and other computer performance modelling
formalisms, these were identified as the source of the error in the
differential equation approximations in~\cite{pepa-odes-ck}. We use
our new \GPEPA\ tool, to explore these switch points and the error
associated with them.

We aim to show that in many cases the existence and location of an
approximation error in the fluid model can be predicted.  Our tool
will be able to warn the modeller about the presence of error for
certain initial parameter regimes or in particular time-intervals for
a given model execution. We examine not only first moment predictions
for several simple performance models but also variances and higher
moment solutions of the ODE approximation. In all cases it is
essential to know when the analysis is accurate. Higher moments, in
particular, can be used to create passage-time analysis
bounds~\cite{pepa-odes-passage} and accurate moment approximations
will make for precise bounds.

\subsection{Motivating Example}
\label{sec:example}

We will first look at a simple motivating example. Consider the
ubiquitous situation of $m$ identical processors running in parallel,
each in need of one of $n$ identical resources. Each processor
repeatedly has to acquire an available resource, after which it is
ready to perform a required task and return to the initial state. Each
resource has to be reset after it is acquired. The actions the system
takes (\eg a processor acquiring a resource or a resource resetting)
are stochastic in nature and do not have a fixed duration. Formally,
this model is defined in \sectref{sec:gpepa}.

\begin{figure}[htb]
\begin{center}
\subfloat[ODE approximation: mean]{\includegraphics[scale=0.60]{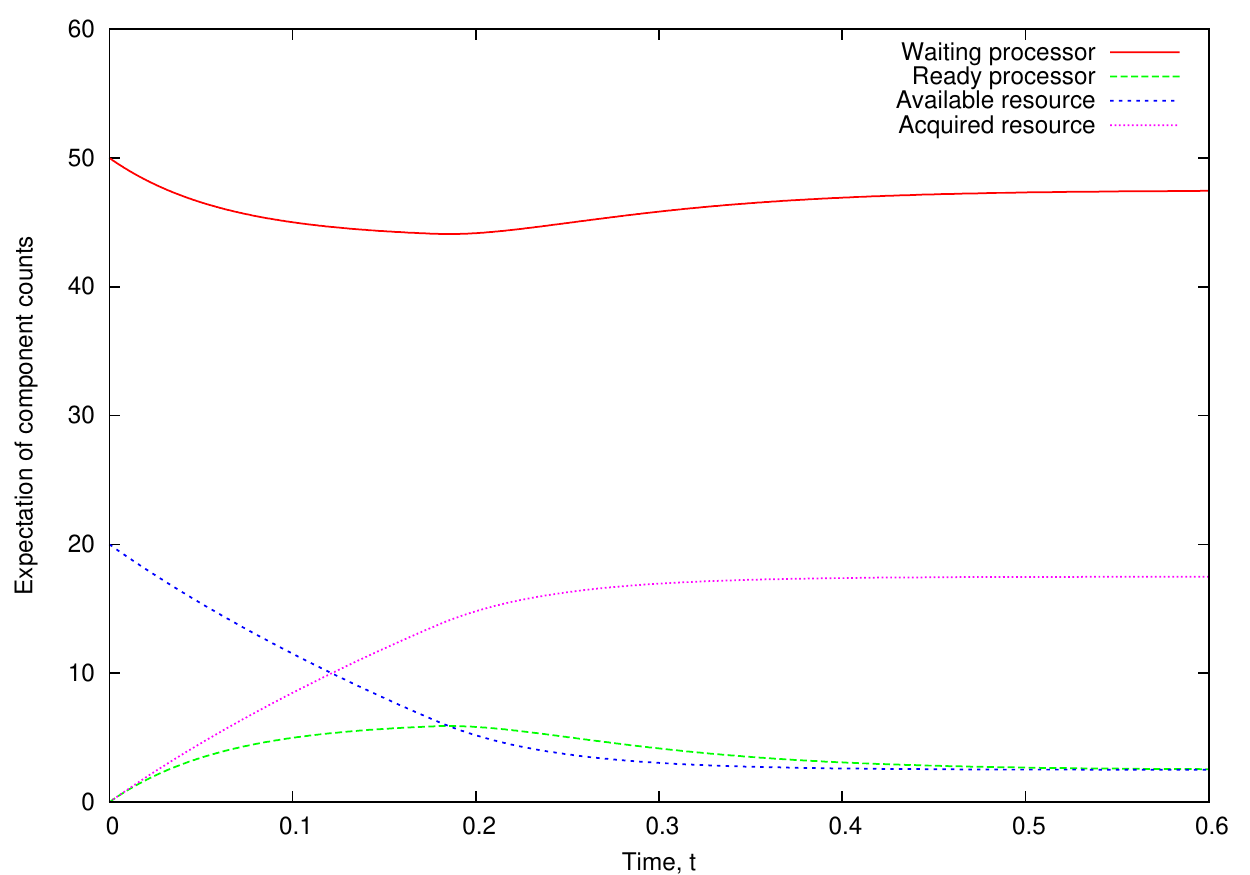}}
\subfloat[Simulation: mean]{\includegraphics[scale=0.60]{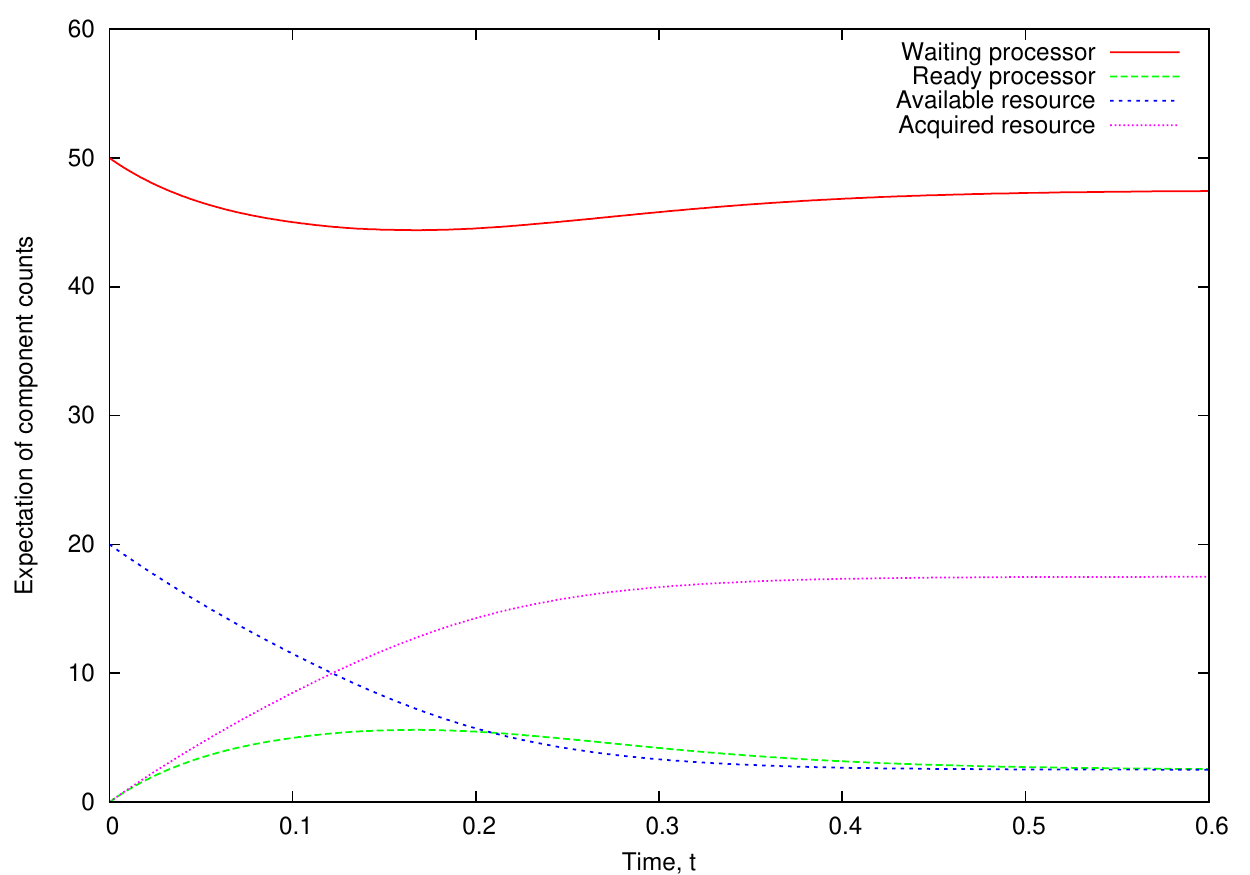}}
\end{center}
\caption{Comparison of the simulation averages and numerical solution to the
corresponding approximating ODEs for the processor/resource model.}
\label{fig:procres_exp}
\end{figure}

We are interested in how the system evolves over time: that is, in
the counts of the four different states within the system ({\em
  available} and {\em acquired} resources and {\em waiting} and {\em
  ready} processors) at time, $t$. One way to do this is to simulate
the system repeatedly and take the mean of counts of each state at
each point of time. An efficient alternative is to deterministically
approximate the evolution of the model by a system of coupled ordinary
differential equations using techniques from, for
example~\cite{pepa-odes-ck,Bor07,Hil05,Car08}. \Figref{fig:procres_exp}
shows a plot for each of these methods for our example, for $m=50$ and
$n=20$.

\begin{figure}[htb]
\begin{center}
  \subfloat[ODE approximation:
  variance]{\includegraphics[scale=0.60]{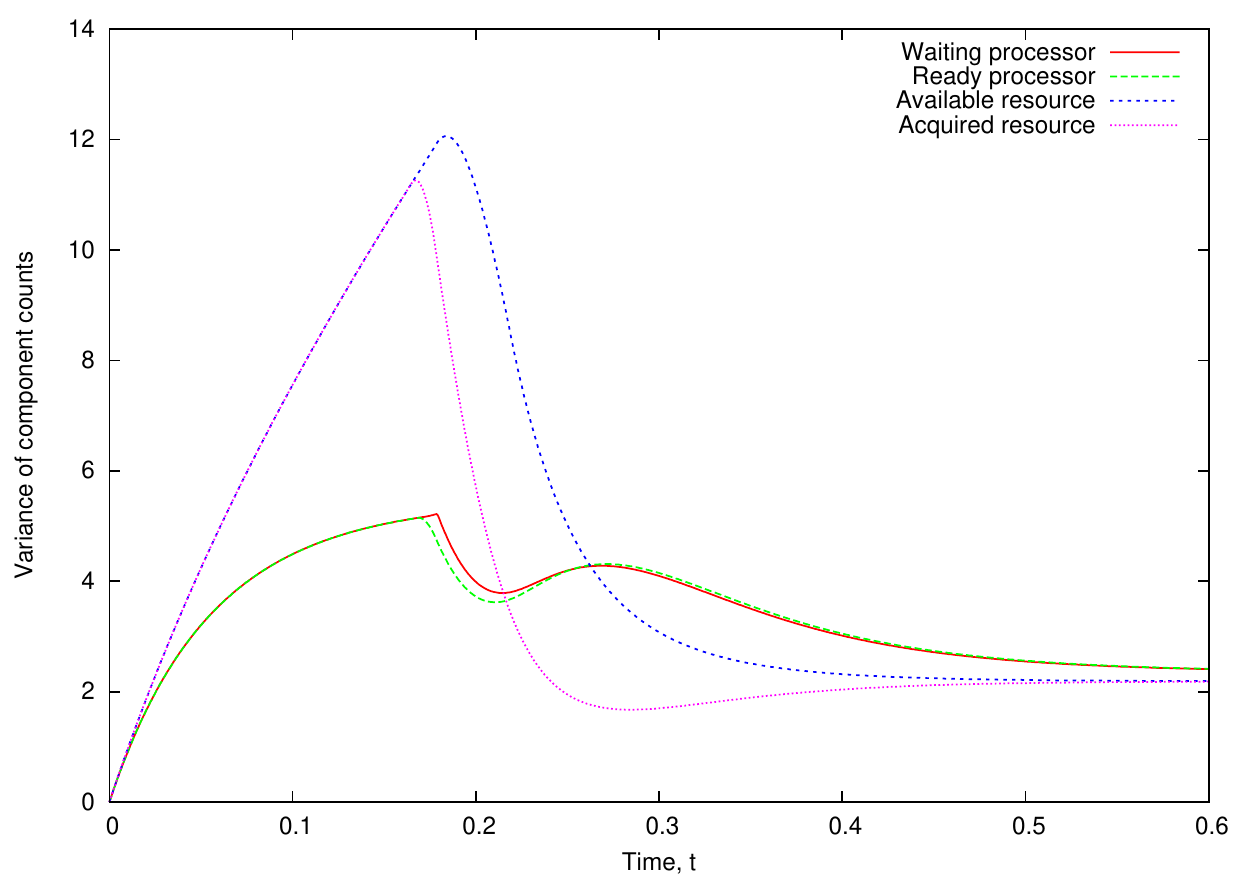}}
  \subfloat[Simulation: variance]{\includegraphics[scale=0.60]{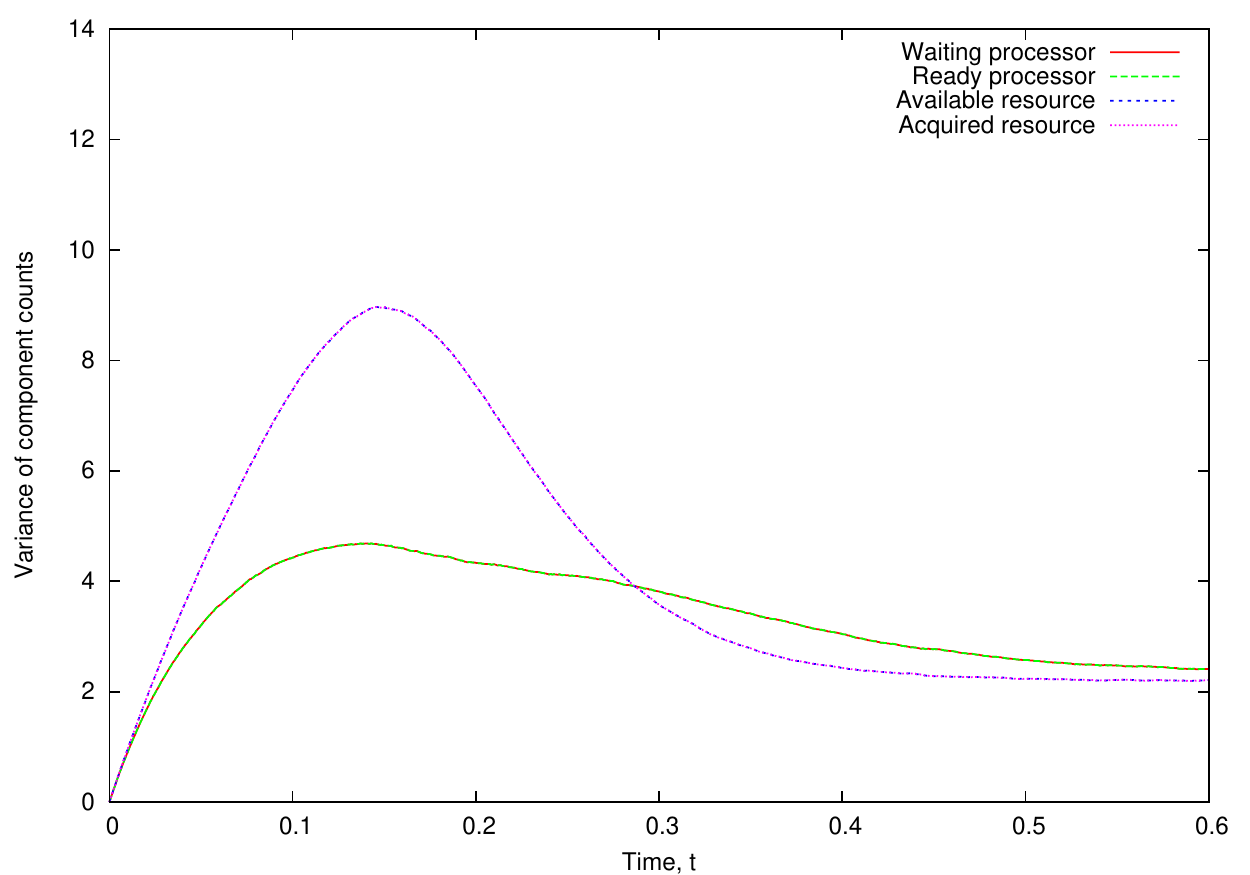}}
\end{center}
\caption{Comparison of the variance for the processor/resource model
obtained from the stochastic simulation with their ODE approximation.}
\label{fig:procres_var}
\end{figure}

It can be useful to know not just the average counts of states at each time but
also the variability of these counts. For the stochastic simulation, this can be
achieved by calculating the variance instead of just the mean. For the ODE
approximation, it has been shown in~\cite{pepa-odes-ck} how to adapt the
existing techniques to produce systems of ODEs approximating higher moments of
state counts. \Figref{fig:procres_var} compares a numerical solution to
these ODEs with the variance obtained from the stochastic simulation.

It is apparent that the ODE approximation is not equally accurate for
the whole observed time. \Figref{fig:procres_exp} and even more so
\figref{fig:procres_var} shows that the error of the approximation
starts just before the time $t=0.2$. 

The main focus of our work is to investigate, with the help of an
efficient software tool, why this error appears and to provide
practical means of detecting it without running the computationally
expensive simulation.  We will explain, using results
from~\cite{pepa-odes-ck}, that it occurs due to models passing through
so-called \emph{switch points}, when the total rate of cooperating
actions between groups of components becomes equal.

\begin{floatingfigure}[r]{8.5cm}
\begin{center}
\includegraphics[scale=0.60]{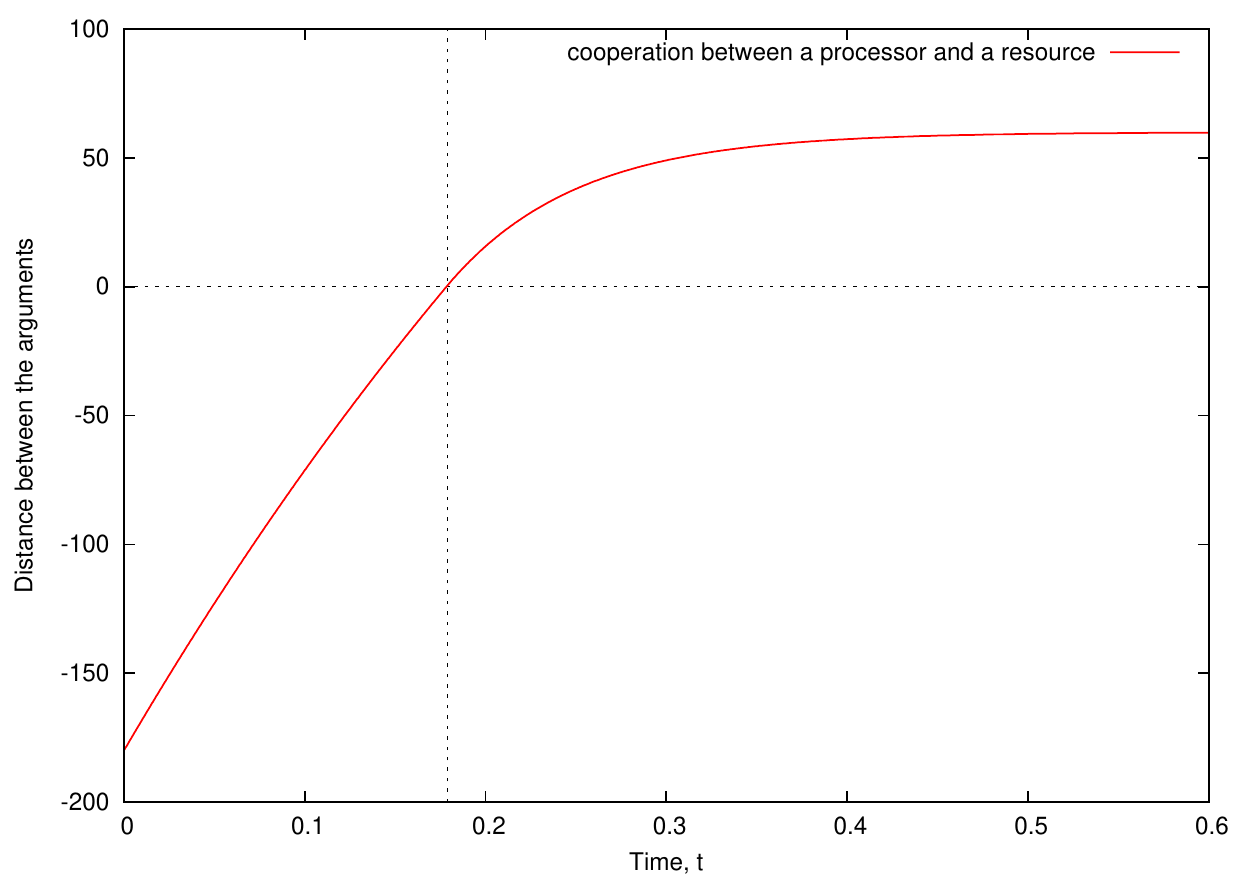}
\end{center}
\caption{Switch point plot showing the difference between total rate
  of processors and resources component groups.
  The switch point occurs when the plot goes through zero.}
\label{fig:procres_switch}
\end{floatingfigure}


We present a tool that will visualise the distance that an evolving
model is from a switch point.  \Figref{fig:procres_switch} shows this
for the processor and resource model and confirms that the error in the
variance approximation coincides with the switch point in total rate
between processors and resources just before $t=0.2$.

In this work, we will consider models described in an extension of the
stochastic process algebra PEPA. \sectref{sec:gpepa} provides a brief
introduction to Grouped PEPA. In \sectref{sec:odes} we will overview
the method of deriving ODE approximations to higher moments of PEPA
models.

It was shown in~\cite{pepa-odes-passage} that the error of this
approximation for the mean decreases as we increase the initial
populations (\eg $m$ and $n$ in this
example). \sectref{sec:theoretical} gives a theoretical justification that a
similar convergence can be expected to hold in the case of the variance ODE
approximation.

It is important to note that although we are studying switch points in
the context of PEPA models, switch points also occur in other
performance modelling settings. Specifically, they occur whenever
there is a contention for a finite pool of resources which may be
saturated by demand.  The most obvious example of this is an $M/M/n$
queue which has $n$ servers, and a service rate of $\mu$ at each
server; the total service rate when $k$ jobs are in the system is
given by $\min(n\mu,k\mu)$, where $k=n$ defines the switch point of
this queue.


\sectref{sec:tool} introduces a tool that for the first time
implements higher order moment approximation for a stochastic process
algebra. This tool is used for our investigations in
\sectref{sec:investigations}, where we present cases studies of
different switch point behaviour and also demonstrate instances of the
results from \sectref{sec:theoretical}.

\subsection{Related Tools}

There are many tools that support analysis of very large state spaces
in performance modelling. Two such popular tools which have good
support for explicit state-space analysis are M\"obius and PRISM.

The M\"obius~\cite{Dea02} framework has
perhaps the widest user base with implementations of many formalisms,
including stochastic process algebras (SPAs), stochastic automata
networks (SANs) and generalised stochastic Petri nets
(GSPNs). M\"obius supports a distributed simulation environment and
numerical solvers for models of up to tens of millions of states.

PRISM~\cite{Kwi02a} is a probabilistic model checker which supports low
level formalisms such as DTMCs, CTMCs and Markov Decision Processes
(MDPs) with an analysis engine based on Binary Decision Diagrams
(BDDs) and Multi-Terminal Binary Decision Diagrams (MTBDDs). PRISM can
analyse models of up to $10^{11}$ states, however this can depend
heavily on the model being studied and on detailed considerations such
as the exact variable ordering in the underlying MTBDD.

Performance tools that support differential-equation based analysis
have been primarily designed around stochastic process algebras such
as stochastic $\pi$-calculus and PEPA. For $\pi$-calculus
SPiM~\cite{Phi04,Phi07} has long been the standard tool for simulating
stochastic $\pi$ calculus models, but being a simulator it suffers
from scalability issues for models with very large populations of
components. A recent tool, JSPiM~\cite{fluid-spatial-pi}, allows for
the \emph{chemical ground form} subset of stochastic $\pi$-calculus to be analysed via
differential equations. For the stochastic process algebra PEPA, the
tools ipc~\cite{Clr07,Brd03e} and the Eclipse PEPA
plug-in~\cite{Tri07} implement the so-called {\em fluid
  translation}~\cite{Hil05} to produce sets of differential equations
for PEPA.

In the field of biological modelling, tools such as Dizzy~\cite{dizzy}
and SPiM have been used to capture first order approximations to
system dynamics using a combination of stochastic
simulation~\cite{gillespie} and differential equation approximation.
A recent tool from~\cite{momentClosureMassAction} generates ODEs
approximating higher moments of models using the mass-action kinetics
and described in the \emph{Systems Biology mark-up language}. However
for the reasons discussed, these techniques do not extend to computer
and communication system modelling.

The tool presented here provides a framework implementation for a
variant of PEPA known as Grouped PEPA or
GPEPA~\cite{passive-fluid-semantics}. Using the established relationship between
the underlying discrete Markov process and the fluid
model~\cite{pepa-odes-ck}, we can for the first time produce higher
moments through differential equation analysis of massive models. The
relationship with the underlying discrete model also allows us to
identify areas of possible inaccuracy where the fluid model is known
to be least accurate. Finally the GPEPA analyser (\GPEPA) allows comparison by
stochastic simulation, where this is feasible.

\subsection{Grouped PEPA models}
\label{sec:gpepa}

\emph{Grouped PEPA} or GPEPA~\cite{pepa-odes-ck} is a simple syntactic
extension of PEPA. A brief summary of PEPA is given in
Appendix~\ref{sec:PEPA}. GPEPA is defined to provide a more elegant
treatment of the ODE moment approximation (and also allowing a more
elegant implementation). Formally, the extension introduces a further
level in the syntax of PEPA, the \emph{Grouped PEPA models}, defined
as:
\begin{equation*}
M  ::= M \sync{L} M ~ \mid \pepagrp{Y}\{P\parallel \cdots \parallel P\}
\end{equation*}
This defines a GPEPA model to be either a PEPA cooperation between
two GPEPA models (over the set of actions $L$) or alternatively a
labelled grouping of PEPA components, $P$, in parallel with each
other. $\pepagrp{Y}$ is the \emph{group label}. The Grouped PEPA model is nothing
more than the standard PEPA model with, additionally, a label to
define the components involved in parallel grouping. These labels will
be later used to define the level at which the ODE approximation to
the system is made.

The example from \sectref{sec:example} can be represented by the
following GPEPA definition: 
\newcommand{\Proc}{\pepa{Processor}}
\newcommand{\Res}{\pepa{Resource}}
\newcommand{\ProcG}{\pepagrp{Processors}}
\newcommand{\ResG}{\pepagrp{Resources}}
\newcommand{\sProc}{\pepa{P}}
\newcommand{\sRes}{\pepa{R}}
\newcommand{\task}{\pepa{task}}
\newcommand{\acquire}{\pepa{acquire}}
\newcommand{\reset}{\pepa{reset}}
\newcommand{\System}{\pepa{System}}
    \begin{align*}
        \Proc_0 & \rmdef (\acquire,\,r_1).\Proc_1&
        \Res_0 & \rmdef (\acquire,\,r_2).\Res_1\\
        \Proc_1 & \rmdef (\task,\,q).\Proc_0 &         \Res_1 & \rmdef (\reset,\,s).\Res_0\\
    \end{align*}
    \vspace{-1cm}
\[        \System \rmdef \pepagrp{Processors}\{ \Proc_0[m] \}
         \sync{\{\acquire\}} \pepagrp{Resources}\{\Res_0[n]\}
\]

The choice of labels ($\pepagrp{Processors}$ and
$\pepagrp{Resources}$) was arbitrary and will serve to identify each
state of the system by counting the occurrences of $\Proc_0$ and
$\Proc_1$ processes in the group $\ProcG$ and occurrences of $\Res_0$
and $\Res_1$ in the group $\ResG$.

\subsubsection{ODE analysis}
\label{sec:odes}
The states in PEPA models originally keep track of the state of each
individual sequential component. This can lead to state space
explosion, which makes the model not amenable to traditional analysis
methods other than the computationally expensive stochastic
simulation. The state space explosion is especially severe (with
respect to the syntactical size of the model) in the case of models
with groups consisting of many components acting in parallel. An
established way to tackle this in the case of groups consisting of
many identical components is by aggregating the state space by keeping
track of \emph{counts} of the individual components~\cite{Hil05}.  In
the context of Grouped PEPA models, it is sufficient to represent each state
of the underlying CTMC by a numerical vector $\mathbf{N}(t)$
consisting of counts $N_{G,P}(t)$ for each possible pair of group
label $G$ and component $P$ (as in~\cite{pepa-odes-ck}).

Allowing this to be a real-valued vector $\mathbf{v}(t)$ (with
components accordingly named $v_{G,P}(t)$), a system of ordinary
differential equations, $\dot{\mathbf{v}}(t) =
\mathbf{f}(\mathbf{v}(t))$ with initial conditions
$\mathbf{v}(0)=\mathbf{N}(0)$ can be intuitively derived from the
corresponding PEPA description. These ODEs deterministically
approximate the PEPA model's evolution over time. This has been
formally shown in~\cite{pepa-odes-ck} (by comparison with the
Chapman--Kolmogorov equations for the underlying CTMC) to be an
approximation to the \emph{expectation} of the individual
group/component counts, \ie$\mathbf{v}(t)\approx
\mathbb{E}[\mathbf{N}(t)]$.

The authors of~\cite{pepa-odes-ck} further extended this method to
derive ODE approximations to higher and joint moments of the
counts, such as variances, covariances and others. The number of differential
equations generated by this method, when calculating all the moments of order up to
$p$, is proportional to $n^p$ where $n$ is the number of component
derivatives.

The method generalises the vectors $\mathbf{N}$ and $\mathbf{v}$ to include
components for values (integers and reals respectively) of each
possible moment of the group/component pair counting processes. For example
$\mathbf{N}(t)$ can include the squared count of component $P$ within
a group $G$, $N_{(P,G)}^2(t)$ and $\mathbf{v}(t)$ an approximation to
its expectation, \ie $v_{(P,G)^2}(t)\approx \mathbb{E}[N_{(P,G)}^2(t)]$.

A general advantage of this approximation to the moments (including
the mean) is that the resulting system of ODEs can be numerically
integrated by standard methods. This usually requires less computation
than running sufficiently many replications of the respective
stochastic simulation over the same model.

\subsubsection{Two-stage Client/Server Example}
\label{sec:clientserver}
\newcommand{\ClientG}{\pepagrp{Clients}}
\newcommand{\ServerG}{\pepagrp{Servers}}
\newcommand{\Client}{\pepa{Client}}
\newcommand{\ClientWaiting}{\pepa{Client}\_\pepa{waiting}}
\newcommand{\ClientThink}{\pepa{Client}\_\pepa{think}}
\newcommand{\Server}{\pepa{Server}}
\newcommand{\ServerGet}{\pepa{Server}\_\pepa{get}}
\newcommand{\ServerBroken}{\pepa{Server}\_\pepa{broken}}
\newcommand{\ClientS}{\pepa{C}}
\newcommand{\ClientWaitingS}{{\pepa{C}_\pepa{w}}}
\newcommand{\ClientThinkS}{{\pepa{C}_\pepa{t}}}
\newcommand{\ServerS}{\pepa{S}}
\newcommand{\ServerGetS}{{\pepa{S}_\pepa{g}}}
\newcommand{\ServerBrokenS}{{\pepa{S}_\pepa{b}}}
\newcommand{\think}{\pepa{think}}
\newcommand{\request}{\pepa{request}}
\newcommand{\data}{\pepa{data}}
\newcommand{\sbreak}{\pepa{break}}
\newcommand{\rcr}{\arp{\pepa{req}}}
\newcommand{\rct}{\arp{\pepa{think}}}
\newcommand{\rsb}{\arp{\pepa{break}}}
\newcommand{\rsd}{\arp{\pepa{data}}}
\newcommand{\rsr}{\arp{\pepa{\reset}}}

We will look at a more complicated example that will be used in the
following investigations. Consider a client/server model, where
communication occurs in two stages. Clients first request a server to
communicate with. After admitting a request from a server, a client
moves to a waiting state and the corresponding server to a ready
state. Then, any ready server can serve a waiting client, after which
the client can perform a required task and the server returns to the
idle state. Additionally, an idle server can break, requiring a
reset. This can be represented by the following PEPA components:
\begin{eqnarray*}
  \Client &\eqdef& (\request,\rcr).\ClientWaiting\\
  \ClientWaiting &\eqdef& (\data,\rsd).\ClientThink\\
  \ClientThink &\eqdef& (\think,\rct).\Client\\[3mm]
  \Server &\eqdef& (\request,\rcr).\ServerGet + (\sbreak,\rsb).\ServerBroken\\
  \ServerGet &\eqdef& (\data,\rsd).\Server\\
  \ServerBroken &\eqdef& (\reset,\rsr).\Server
\end{eqnarray*}
composed into the GPEPA system equation defined by $CS(c,s)$ over the
set of actions $L = \{\request,\data\}$:
\begin{equation*}
CS(c,s) = \ClientG\{\Client[c]\}\sync{L}\ServerG\{\Server[s]\}
\end{equation*}
The aggregation described above represents each derivative state of
the model at each time $t$ as a vector $\mathbf{N}(t)$, \ie the
underlying CTMC can be treated as a vector valued stochastic process,
with components $N_{\ClientG,P}(t)$, where $P$ can be $\Client$,
$\ClientWaiting$ or $\ClientThink$ and $N_{\ServerG,Q}(t)$ where $Q$
can be $\Server$, $\ServerGet$ or $\ServerBroken$. The ODE
approximation generates a system of ODEs, $\dot{\mathbf{v}}(t) =
\mathbf{f}(\mathbf{v}(t))$ with a real-valued vector solution
$\mathbf{v}(t)$ with the same components as the vector $\mathbf{N}(t)$
and initial conditions
$v_{\ClientG,\Client}(0)=N_{\ClientG,\Client}(0) = c$,
$v_{\ServerG,\Server}(0)=N_{\ServerG,\Server}(0) = s$ and $v_{G,P}(0)
= 0$ otherwise.  See \sectref{sec:theoretical} for the exact formulation of
$\mathbf{f}(\mathbf{v}(t))$.

\subsection{Nature of the approximations}
\label{sec:approx}
The nature of the approximation of the system of ODEs to the CTMC
evolution boils down to the approximation of the rate at which two
components evolve in cooperation (as discussed originally
in~\cite{pepa-odes-ck}).

This is trivial in the unsurprising case of \emph{purely concurrent models},
where no cooperation takes part. For these, the approximation is exact and can
directly replace moment calculation via stochastic simulation. 
\label{sec:switch}

In general, the rate of cooperation can be of the form
$\mathbf{r}(\mathbf{N}(t))\cdot
\min(\mathbf{f}(\mathbf{N}(t)),\mathbf{g}(\mathbf{N}(t)))$ where
$\mathbf{r}(\cdot)$ is a rational function of piecewise linear
functions and $\mathbf{f}(\cdot)$ and $\mathbf{g}(\cdot)$ are
piecewise linear functions all of the components of
$\mathbf{N}(t)$. The functions $\mathbf{f}(\mathbf{N}(t))$ and
$\mathbf{g}(\mathbf{N}(t))$ are a calculation of the total rates of
cooperating actions being enabled in cooperating component
groups.

For a certain class of models called the \emph{split-free} models, introduced in
\cite{pepa-odes-ck}, it
can be shown that the function $\mathbf{r}$ is constant and
$\mathbf{f},\mathbf{g}$ are piecewise linear.  It then turns out that
the nature of the deterministic approximation of $\mathbf{N}(t)$ by
$\mathbf{v}(t)$ depends on the approximation
\begin{equation}
 \mathbb{E}[\min(\mathbf{f}(\mathbf{N}(t)),\mathbf{g}(\mathbf{N}(t))] \approx
 \min(\mathbf{f}(\mathbb{E}[\mathbf{N}(t)]),\mathbf{g}(\mathbb{E}[\mathbf{N}(t)]))
\label{eqn:switch_approx}
\end{equation}
where the right hand side is approximated by
$\min(\mathbf{f}(\mathbf{v}(t)),\mathbf{g}(\mathbf{v}(t)))$. Note that
the functions $\mathbf{f}$ and $\mathbf{g}$ may also include further
minimum terms themselves and thus induce multiple further applications
of the approximation not shown explicitly above. It is argued
in~\cite{pepa-odes-ck} that the error of this approximation is at its
highest around 
so-called \emph{switch points}.  These occur when the total rate of
cooperating action between component groups becomes equal and hence
$\mathbf{f}(\mathbf{N}(t))=\mathbf{g}(\mathbf{N}(t))$ causing the
$\min$ function to switch.

One of the main contributions of this paper is to verify and further
investigate this claim empirically. We will be able to use our GPEPA
tool to identify the switch points and to display the error in the ODE
approximation around the switch points of a given GPEPA model. We will
focus on \emph{first order} switch points: those coming from a
$\min$ term involving no higher orders than means of component counts.

The switch points are also defined for general GPEPA models. However,
for those, the nature of the ODE moment approximation also depends on
the approximation of terms, $\mathbb{E}[\mathbf{r}(\mathbf{N}(t))]$
for a rational function $\mathbf{r}(\cdot)$. Our GPEPA tool allows
analysis of these models, by using the approximation
$\mathbf{r}(\mathbb{E}[\mathbf{N}(t)])$. However, in our first investigation we
concentrate on the split-free models. We will be using the
client/server model $CS(c,s)$ from the above section, as it can be
easily shown to be split-free.



\subsection{Theoretical considerations}
\label{sec:theoretical}
In this section, we aim to provide justification for the convergence of the
suitably-scaled mean and variance ODE approximations as the component
population sizes are increased.

In order to do this, we will more formally set up some notation to allow us to
consider general GPEPA models and their associated systems of ODEs in a
compact manner. For a general GPEPA model, assume that we have a vector-valued
stochastic process $\mathbf{N}(t)$, defined on $\realnn$ taking values in
$\intnn^N \subset \realnn^N$, for some $N \in \intnn$. In line with
\sectref{sec:odes}, each component of this process counts the number of a
particular derivative state currently active in a parallel group of the model,
of which there are $N$ derivative states in total, across all parallel groups.

Analogously, we define the vector-valued deterministic function
$\mathbf{v}(t)$, also defined on $\realnn$ and taking values in $\realnn^N$ to
be the ODE approximation to the expectation of this model. We assume that it
is defined uniquely by the system of differential equations $\dot{\mathbf{v}}(t)
= \mathbf{f}(\mathbf{v}(t))$
and the initial condition $\mathbf{v}(0) = \mathbf{N}(0)$.  As discussed above,
the exact form of the deterministic function $\mathbf{f} : \realnn^N
\rightarrow \real^N$ is given in~\cite{pepa-odes-ck} and corresponds
component-wise to the rate at which each derivative state is incremented, minus
that at which it is decremented, in a given state of the model. It helps now to
consider the system of ODEs explicitly for an example model. Indeed, in the
case of the client/server example of \sectref{sec:clientserver}, we have a total
of six derivative states, giving $N=6$,
\begin{equation*}
\begin{split}
\mathbf{N}(t) &\equiv
(N_C(t),\,N_{C_w}(t),\,N_{C_t}(t),\,N_S(t),\,N_{S_g}(t),\,N_{S_b}(t))^\text{T}\\
\mathbf{v}(t) &\equiv
(v_C(t),\,v_{C_w}(t),\,v_{C_t}(t),\,v_S(t),\,v_{S_g}(t),\,v_{S_b}(t))^\text{T}
\end{split}
\end{equation*}
and:
\begin{equation*}
\mathbf{f}(\mathbf{v}(t)) \equiv 
\begin{pmatrix} 
-\min(v_S(t),\,v_C(t)) \rcr + v_{C_t}(t) \rct\\
-\min(v_{C_w}(t),\,v_{S_g}(t)) \rsd + \min(v_S(t),\,v_C(t)) \rcr\\
-v_{C_t}(t) \rct + \min(v_{C_w}(t),\,v_{S_g}(t)) \rsd\\
-\min(v_S(t),\,v_C(t)) \rcr - v_S(t) \rsb + \min(v_{C_w}(t),\,v_{S_g}(t)) \rsd +
v_{S_b}(t) \rsr\\
-\min(v_{C_w}(t),\,v_{S_g}(t)) \rsd + \min(v_S(t),\,v_C(t)) \rcr\\
-v_{S_b}(t) \rsr + v_S(t) \rsb
\end{pmatrix}
\end{equation*}
where we have used an appropriate shorthand notation for the components of
$\mathbf{N}(t)$ and $\mathbf{v}(t)$.  In order to make the desired theoretical
considerations in this section, it is necessary to consider a
\emph{structurally equivalent} sequence of GPEPA models which have the same
structure of parallel component groups and differ only in terms of the size of
the component populations within these groups.  Furthermore, we require that
they all have the same initial proportion of each component type in each case.
For such a sequence of GPEPA models, let $\{\mathbf{N}^i(t)\}_{i=1}^\infinity$
be their associated stochastic counting processes in the notation above, and
for each $i$, write $S_i := N^i_1(t) + \cdots + N^i_n(t)$ for the total
component population of model $i$.\footnote{This does not depend on $t$ because
the PEPA operational semantics preserve component populations.} So our
requirement of constant initial component type proportions is stated formally
as:
\begin{equation*}
\frac{N^i_k(0)}{S_i} = \frac{N^j_k(0)}{S_j} \text{ for all $i,\,j > 0$ and $k \in
\{1,\,\ldots,\,n\}$}
\end{equation*}
In the case of (G)PEPA, it is relatively straightforward to see that for any
$\mathbf{x} \in \realnn^N$, $\mathbf{f}(k \mathbf{x}) = k
\mathbf{f}(\mathbf{x})$ for all $k \in \realnn$. Furthermore, since the GPEPA
models in our sequence differ only in terms of their initial component counts,
it is easy to see that the function $\mathbf{f}(\cdot)$ is the same for any
$i$. These two facts together mean that we need only define the fluid
approximation to $\mathbf{N}^i(t)$, say, $\mathbf{v}^i(t)$ for a particular
value of $i$, and the fluid approximation for any other $i$ can be defined in
terms of it. Indeed, for any $i,\,j > 0$, if $\dot{\mathbf{v}}^i(t) =
\mathbf{f}(\mathbf{v}^i(t))$ and $\dot{\mathbf{v}}^j(t) =
\mathbf{f}(\mathbf{v}^j(t))$ with initial conditions, $\mathbf{v}^i(0) =
\mathbf{N}^i(0)$ and $\mathbf{v}^j(0) = \mathbf{N}^j(0)$, respectively, we
have:
\begin{equation*}
\mathbf{v}^i(0) = \mathbf{v}^j(0) \times \frac{S_i}{S_j} \text{ and }
\mathbf{v}^i(t) = \mathbf{v}^j(t) \times \frac{S_i}{S_j} \text{ for all $t > 0$ }
\end{equation*}
Thus for the rest of this section, we consider the quantity
$\mathbf{\bar{v}}(t) := \mathbf{v}^i(t) / S_i$, for all $t > 0$, which we have
just seen is independent of $i$.

\subsubsection{Convergence of mean approximation}

It is shown in~\cite{pepa-odes-passage}, based on a result by Kurtz
\cite{kurtzjump}, that, in the above notation for any
sequence of structurally equivalent GPEPA models, the following holds:

\begin{theorem}
If $S_i \rightarrow \infinity$ as $i \rightarrow \infinity$ then: 
\begin{equation*}
\frac{1}{S_i} \mathbb{E}[\mathbf{N}^i(t)] \rightarrow \mathbf{\bar{v}}(t)
\end{equation*}
as $i
\rightarrow \infinity$, uniformly over bounded intervals $t \in [0,\,T)$.
\end{theorem} 

This demonstrates that, in the limit of component population size, the mean
number of components at time $t$ does indeed tend towards the ODE solution. An
example of this convergence is shown in \figref{fig:twostage_error}(a), \sectref{sec:investigations}. 


\subsubsection{Convergence of variance approximation}
\label{sec:conv_var}

In this section, we provide theoretical support for our hypothesis that in the
limit of large populations and under an appropriate scaling, the variance of
component counts converges to the approximation given by the ODEs for
split-free GPEPA models.\footnote{Similar considerations should also be possible
for non split-free models, but we do not consider these here for brevity.}

We will start by approximating the GPEPA model's underlying CTMC, $\mathbf{N}^i(t)$, as the sum of a deterministic process, $\mathbf{\bar{v}}(t)$, given by the
first order ODEs, and a Gaussian
process, $\mathbf{E}(t)$ defined below. From this description, we will derive a set of ODEs
describing the evolution of the covariances of the Gaussian process. These can
be formally shown (\thrmref{t:fclt}) to agree with the covariances of the CTMC in the limit of a sequence of GPEPA
models of increasing size. We will argue that the second moment ODEs from the
CTMC (\sectref{sec:odes}) capture the dynamics of the system more accurately than the ODEs generated from the Gaussian process. This provides the basis
for our hypothesis that the variances of the CTMC do indeed converge to the solution to the second moment ODEs
from \sectref{sec:odes} as the population size increases.

The decomposition described above gives the following approximation to the underlying CTMC of a GPEPA model:
\begin{equation}
\label{e:s-order-approx}
\mathbf{N}^i(t) \approx S_i \mathbf{\bar{v}}(t) + \sqrt{S_i} \mathbf{E}(t)
\end{equation}
where $\mathbf{E}(t)$ is the Gaussian process mentioned above.
In order to proceed with defining $\mathbf{E}(t)$, it is necessary to
enumerate explicitly the transitions in the aggregated state space of the
GPEPA model. Assume that there are $K$ such transitions and,
following~\cite{pepa-fclt}, let $\{\mathbf{l}^k \in \ints^N\}_{k=1}^K$ be the
sequence of jump vectors, specifying that if the $k$th such transition occurs
at time $t$, $\mathbf{N}(t) = \mathbf{N}(t-) + \mathbf{l}^k$, where $t-$ is the
instant immediately preceding $t$. Then define rate
functions, $\{f^k : \realnn^N \rightarrow \realnn\}_{k=1}^K$, specifying the
aggregate rate of each transition. In the case of the client/server example,
we have $K = 5$ (for the $5$ possible activities) and:
\begin{equation*}
  \begin{array}{lll}
    \mathbf{l}^1 = (-1,\,1,\,0,\,-1,\,1,\,0)  &\qquad & f^1(\mathbf{x}) = \min(x_1,\,x_4)\, \rcr\\
    \mathbf{l}^2 = (0,\,-1,\,1,\,1,\,-1,\,0)  &\qquad & f^2(\mathbf{x}) = \min(x_2,\,x_5)\, \rsd\\
    \mathbf{l}^3 = (1,\,0,\,-1,\,0,\,0,\,0)  &\qquad &  f^3(\mathbf{x}) = x_3\, \rct\\
    \mathbf{l}^4 = (0,\,0,\,0,\,-1,\,0,\,1)  &\qquad &  f^4(\mathbf{x}) = x_4\, \rsb\\
    \mathbf{l}^5 = (0,\,0,\,0,\,1,\,0,\,-1)  &\qquad &  f^5(\mathbf{x}) = x_6\, \rsr
  \end{array}
\end{equation*}
The stochastic process $\mathbf{E}(t)$ is now given in the following
theorem~\cite{pepa-fclt}, which applies to any sequence of
structurally equivalent split-free GPEPA models with underlying CTMCs
$\{\mathbf{N}_i(t)\}_{i=1}^\infinity$, corresponding model sizes
$\{S_i\}_{i=1}^\infinity$ and rescaled ODE approximation
$\mathbf{\bar{v}}(t)$.

\thrmref{t:fclt} below defines the Gaussian process $\mathbf{E}(t)$ in terms of
time-scaled Wiener processes, $W^k(t)$. Readers unfamiliar with Wiener processes
and weak convergence of stochastic processes can consult for instance Rogers and
Williams~\cite{diff-mark-and-mart} or Klebaner~\cite{intro-stoch-calc}.  The
set $\hat{\mathbf{T}}$ defined in the theorem can be given more intuitively as
the set of all times at which two arguments of a minimum function in the
right-hand side, $\mathbf{f}(\mathbf{\bar{v}}(t))$, of the first moment ODEs are
equal.

\begin{theorem}
\label{t:fclt}
Let $T > 0$ and let $\hat{\mathbf{T}}$ be the subset of $\{t \in [0,\,T)\}$
for which $\mathbf{f}(\cdot)$ is not totally differentiable at the point
$\mathbf{\bar{v}}(t)$.  We require that $\hat{\mathbf{T}}$ has Lebesgue measure
zero.  Then on all of $[0,\,T) \setminus \hat{\mathbf{T}}$,
$\mathbf{f}(\cdot)$ has a well-defined Jacobian at the point $\mathbf{\bar{v}}(t)$,
say $D\,\mathbf{f}(\mathbf{\bar{v}}(t))$. Extend this to all points
$\{\mathbf{\bar{v}}(t)\,:\,t \in [0,\,T)\}$, say by defining it to be the matrix of
zeros at times in $\hat{\mathbf{T}}$.

Then if $S_i \rightarrow \infinity$ as $i \rightarrow \infinity$,
$\frac{\mathbf{N}^i(t)}{\sqrt{S_i}} - \sqrt{S_i}\mathbf{\bar{v}}(t)
\Rightarrow \mathbf{E}(t)$ also as $i \rightarrow \infinity$, where:
\begin{equation*}
\mathbf{E}(t) := \int_0^t D\,\mathbf{f}(\mathbf{\bar{v}}(s)) \cdot
\mathbf{E}(s) \,{\rm d}s + \sum_{k \in K} W^k\left(\int_0^t f^k(\mathbf{\bar{v}}(s))\,{\rm d}s\right) \mathbf{l}^k
\end{equation*}
and $\{W^k(t)\}_{k=1}^K$ is a sequence of $K$ mutually independent standard
Wiener processes (aka Brownian motions). The convergence ($\Rightarrow$) is weak convergence in $D_{\realnn^N}[0,\,T)$, the space of $\realnn^N$-valued
c\`adl\`ag\footnote{Continue \`a droite, limit\'ee \`a gauche, that is, right
continuous with left limits.} functions.
\end{theorem}
This is essentially a generalisation of a result of
Kurtz~\cite{kurtz-strong-approx-mcs} to cope with the case of non-smooth rate
functions as introduced in the case of PEPA models, or other formalisms for
modelling synchronisation in computer systems, by the minimum functions.  It can
be shown that the process $\mathbf{E}(t)$ is the unique solution
of the following (It\={o}) stochastic differential equation (SDE):
\begin{equation*}
{\rm d} \mathbf{E}(t) = \mathbf{\mu}(\mathbf{E}(t),\,t)\,{\rm d}t + 
\mathbf{\sigma}(t)\, {\rm d}\mathbf{W}(t)
\end{equation*} 
where $\mathbf{\mu}(\mathbf{x},\,t) : \real^N \times \realnn \rightarrow
\real^N$ and $\mathbf{\sigma}(t) : \realnn
\rightarrow \real^{N\times K}$ are defined by:
\begin{equation*}
\mathbf{\mu}(\mathbf{x},\,t) := D\,\mathbf{f}(\mathbf{\bar{v}}(t)) \cdot
\mathbf{x},\qquad
\mathbf{\sigma}(t) := \left( l^j_i \times \sqrt{f^j(\mathbf{\bar{v}}(t))} \right)_{ij}
\end{equation*}
and $\mathbf{W}(t) := (W^1(t),\,\ldots,\,W^K(t))^\text{T}$ is a $K$-dimensional
standard Wiener process. This representation allows us to apply the machinery of
It\={o}'s Lemma, see for example \cite{diff-mark-and-mart,foundations-modern-prob}, to derive the following system of ODEs, whose unique
solution is exactly the covariance matrix of $\mathbf{E}(t)$:
\begin{align}
\label{e:e-covar-odes}
\frac{\rm d}{{\rm d}t} \mathbf{Cov}[\mathbf{E}(t),\,\mathbf{E}(t)] = {} &
\mathbf{Cov}[\mathbf{E}(t),\,\mathbf{E}(t)] \,
(D\,\mathbf{f}(\mathbf{\bar{v}}(t)))^\text{T} + 
D\,\mathbf{f}(\mathbf{\bar{v}}(t)) \,
\mathbf{Cov}[\mathbf{E}(t),\,\mathbf{E}(t)]^\text{T}
 + \sum_{k \in K} f^k(\mathbf{\bar{v}}(t)) \, \mathbf{l}^k \,
(\mathbf{l}^k)^\text{T}
\end{align}
If we apply this to the client/server example for the specific
covariance $E_S(t)$, corresponding to the $\Server$ component of
$\mathbf{N}(t)$, $N_S(t)$, we have:
\begin{align}
\label{e:fclt-var-ode}
\frac{\rm d}{{\rm d}t} \text{Cov}[E_S(t),\,E_S(t)] = {} &
- 2 \, \rcr \left( \mathbf{1}_{\{\bar{v}_S(t) \leq \bar{v}_C(t)\}}
\text{Cov}[E_S(t),\,E_S(t)]
+ \mathbf{1}_{\{\bar{v}_S(t) > \bar{v}_C(t)\}}
\text{Cov}[E_S(t),\,E_C(t)] \right)\nonumber\\
& + 2 \, \rsd \left( \mathbf{1}_{\{\bar{v}_{S_g}(t) \leq \bar{v}_{C_w}(t)\}}
\text{Cov}[E_S(t),\,E_{S_g}(t)] +
\mathbf{1}_{\{\bar{v}_{S_g}(t) > \bar{v}_{C_w}(t)\}}
\text{Cov}[E_S(t),\,E_{C_w}(t)] \right)\nonumber\\
& - 2 \, \rsb \text{Cov}[E_S(t),\,E_S(t)]
+ 2 \, \rsr \text{Cov}[E_S(t),\,E_{S_b}(t)]\nonumber\\
& + \rcr \min(\bar{v}_C(t),\,\bar{v}_S(t)) +
\rsd \min(\bar{v}_{C_w}(t),\,\bar{v}_{S_g}(t)) + 
\rsb \bar{v}_S(t) + \rsr \bar{v}_{S_b}(t)
\end{align}
Note that \thrmref{t:fclt} suggests the approximation:
\begin{equation*}
\mathbf{Cov}[N_S(t),\,N_S(t)] \approx \mathbf{Cov}[S \bar{v}_S(t) + \sqrt{S}
E_S(t),\,S \bar{v}_S(t) + \sqrt{S} E_S(t)] = S \,
\mathbf{Cov}[E_S(t),\,E_S(t)]
\end{equation*}
where $S$ is taken as the total population of clients and servers. Indeed, in
general, \thrmref{t:fclt} implies (assuming its hypothesis), as $i \rightarrow \infinity$, if $S_i
\rightarrow \infinity$, that:
\begin{equation*}
\frac{1}{S_i} \, \mathbf{Cov}[\mathbf{N}^i(t),\,\mathbf{N}^i(t)] \rightarrow
\mathbf{Cov}[\mathbf{E}(t),\,\mathbf{E}(t)]
\end{equation*}
So the system of ODEs given in \eqnref{e:e-covar-odes} yields an approximation
to the covariance matrix of the underlying CTMC of a general GPEPA
model. Furthermore, as we have just seen, this approximation is guaranteed by
\thrmref{t:fclt} to converge in the limit of large populations.

However, the covariance ODE approximation implemented by the \GPEPA\ tool
consists of integrating a slightly different system of ODEs. Our intention now
is to show that they are very similar to those of \eqnref{e:e-covar-odes} and
in fact can intuitively be regarded as a better approximation to the actual
covariance matrix of the CTMC. This is the basis of our conjecture that a similar
convergence result also holds, and furthermore, that the rate of convergence
may well be faster for the ODEs implemented in the \GPEPA\ tool.

In more detail for our example, applying the techniques
of~\cite{pepa-odes-ck} for the specific term $\text{Cov}[N_S(t),\,N_S(t)]$,
we obtain the exact differential equation:
\begin{align}
\label{e:ctmc-var-ode}
\frac{\rm d}{{\rm d}t} \text{Cov}[N_S(t),\,N_S(t)] = {} &
\frac{\rm d}{{\rm d}t} \mathbb{E}[N_S^2(t)] - 2 \mathbb{E}[N_S(t)] \frac{\rm d}{{\rm d}t}
\mathbb{E}[N_S(t)]\nonumber\\
= {} &
- 2 \, \rcr \biggl( \mathbb{E}[\min(N_C(t)\,N_S(t),\,N_S^2(t))] -
\mathbb{E}[\min(N_C(t),\,N_S(t))] \mathbb{E}[N_S(t)] \biggl)\nonumber\\
& + 2 \, \rsd \biggl( \mathbb{E}[\min(N_{C_w}(t)\,N_S(t),\,N_{S_g}\,N_S(t))]
- \mathbb{E}[\min(N_{C_w}(t),\,N_{S_g}(t))] \mathbb{E}[N_S(t)] \biggl)\nonumber\\
& - 2 \, \rsb \biggl( \mathbb{E}[N_S^2(t)] - \mathbb{E}^2[N_S(t)] \biggl)\nonumber\\
& + 2 \, \rsr \biggl( \mathbb{E}[N_{S_b}(t)\,N_S(t)] - \mathbb{E}[N_{S_b}(t)]\,\mathbb{E}[N_S(t)]
\biggl)\nonumber\\
& + \rcr \, \mathbb{E}[\min(N_C(t),\,N_S(t))]
+ \rsd \, \mathbb{E}[\min(N_{C_w}(t),\,N_{S_g}(t))]\nonumber\\
& + \rsb \, \mathbb{E}[N_S(t)]
+ \rsr \, \mathbb{E}[N_{S_b}(t)] 
\end{align}
Since the corresponding system of ODEs cannot be solved analytically or
numerically due to the presence of expectations of non-linear functions, the
approximation $\mathbb{E}[\min(X,\,Y)] \approx
\min(\mathbb{E}[X],\,\mathbb{E}[Y])$ can be applied repeatedly as
in~\cite{pepa-odes-ck} and~\sectref{sec:approx} to deduce approximate ODEs for
$\text{Cov}[N_S(t),\,N_S(t)]$ and the other covariances. This system of ODEs
can then be solved numerically as implemented in general by the \GPEPA\ tool.

There are two kinds of approximations being applied here, those which we might
call \emph{first-order}, such as $\mathbb{E}[\min(N_C(t),\,N_S(t))] \approx
\min(\mathbb{E}[N_C(t)],\,\mathbb{E}[N_S(t)])$, and those which we might call
\emph{second-order}, such as $\mathbb{E}[\min(N_C(t)\,N_S(t),\,N_S^2(t))]
\approx \min(\mathbb{E}[N_C(t)\,N_S(t)],\,\mathbb{E}[N_S^2(t)])$. The key
point to note now is that if we keep the first-order approximations the same but
replace the second-order approximations by ones of the form:
\begin{equation*}
\mathbb{E}[\min(N_C(t)\,N_S(t),\,N_S^2(t))] \approx
\mathbf{1}_{\{\mathbb{E}[N_S(t)] > \mathbb{E}[N_C(t)]\}} \,
\mathbb{E}[N_C(t)\,N_S(t))] + \mathbf{1}_{\{\mathbb{E}[N_S(t)] \leq
\mathbb{E}[N_C(t)]\}} \, \mathbb{E}[N_S^2(t)]
\end{equation*}
then we recover the system of ODEs of \eqnref{e:e-covar-odes}. This is a
reasonable approximation, but we switch second-order minimum terms by
making only first-order comparisons. It is thus intuitively clear that it is
likely to be a worse approximation than the second moment ODEs derived from the
CTMC.

On this basis, we might reasonably expect that the covariance ODE
approximations, derived from the CTMC and implemented by the GPA tool, should
converge to the actual covariances when scaled by the total component population
size. An example of this can be seen in \figref{fig:twostage_error}(b),
\sectref{sec:investigations}.

\section{\GPEPA: The GPEPA Analyser}
\label{sec:tool}
In this section we introduce a new tool for analysing Grouped PEPA models. The
\emph{Grouped PEPA analyser} (\GPEPA) uses the results from~\cite{pepa-odes-ck}, briefly described above, as a basis for producing deterministic approximations of transient behaviour of
syntactically specified Grouped PEPA models. It can also use these to provide passage-time
approximations from~\cite{pepa-odes-passage}.

\GPEPA\ in addition implements stochastic simulation of the models to
allow investigations into the nature of the approximation and the
convergence results from \sectref{sec:theoretical}. In the next
section, we look at some specific examples that empirically
demonstrate this.  We first give an overview of \GPEPA, giving its
syntax and the commands it provides.

\subsection{Overview}
The syntax of models specified in \GPEPA\ is very close to the formal
language of Grouped PEPA. Each \GPEPA\ model starts with definitions
of parameters used as rates in component definitions and
parameters used as initial counts in the grouped model
definitions. The definitions of individual PEPA components then
follow. Finally, a
single Grouped PEPA model is given, using the defined components.

Groups are specified by labels with components enclosed in braces
\texttt{\{} \texttt{\}}. Multiple components of the same type are
given by \texttt{[n]} written after the component identifier, where
\texttt{n} is a previously-defined parameter. The cooperation
operator is represented by the cooperation set  enclosed in angled
brackets \texttt{<} \texttt{>}.

The GPEPA model corresponding to the example from \sectref{sec:example} can be
represented in \GPEPA\ as:
\begin{code}
r1=2.0; q=14.0;  m=50.0;                r2=14.0; s=2.0;  n=20.0;
            
Processor0 = (acquire,r1).Processor1;   Resource0 = (acquire,r2).Resource1;
Processor1 = (task,q).Processor0;       Resource1 = (reset,s).Resource0;

      Processors{Processor0[m]}<acquire>Resources{Resource0[n]}
\end{code}

\appdxref{appendix:gpepa_syntax} contains the grammar describing the full syntax of \GPEPA.

\subsection{ODE Analysis and Comparison with Simulation}
On each \GPEPA\ model, several analyses can be performed. Each
analysis is specified with required parameters (for example, the time
range we are interested in for the transient behaviour of the model),
after the analysis name. Following this is a list of commands that
allow visualisation and further manipulation of the resulting data
points.

The \emph{ODE analysis} provides analysis of the given model by a deterministic
approximation via a set of ordinary differential equations as described in
\sectref{sec:odes}. The enclosed commands determine the maximum order that has to
be considered in calculations. For example, if the commands involve only plots
of means and variances of component counts, differential equations for approximation of
all the possible (joint) moments of order $1$ and $2$ are implicitly generated. These ODEs are then
numerically integrated (using a fourth-order Runge--Kutta solver). The parameter
\texttt{stopTime}
specifies the time over which the numerical solution is given. Parameter
\texttt{stepSize} determines the fixed time interval between each successive pair of data points that are taken. The parameter \texttt{density} specifies the
accuracy by telling the solver how many sub-intervals should each time interval
between the data points be divided into. The following \GPEPA\ code performs an
ODE analysis on the above \GPEPA\ model, displaying: mean counts of $\Proc_0$ components in group $\ProcG$ and $\Res_0$ components in group $\ResG$ and model switch points:
\begin{code}
odes(stopTime = 3.0,stepSize = 0.001,density=10){
     plot(E[Processors:Processor0],E[Resources:Resource0]); 
     plotSwitchpoints(1); ...
}
\end{code}

The \emph{Simulation analysis} provides analysis of models by
stochastic simulation of their underlying continuous time Markov
chain. \GPEPA\ generates a representation of the CTMC and uses the
Gillespie algorithm~\cite{gillespie} to simulate the CTMC at given
time intervals until the simulation stop time is reached. 

The stochastic simulation is repeated several times (given by the
parameter \texttt{replications}) and then averaged to provide the
final data set. On the above model, the following code performs a
simulation analysis to extract the same mean component counts as before:
\begin{code}
simulation(stopTime = 3.0,stepSize = 0.001,replications=1000){
     plot(E[Processors:Processor0],E[Resources:Resource0]); ...
}
\end{code}

To compare results from stochastic simulation and ODE analysis, the
\emph{Comparison} analysis provides a way of calculating the difference between
the two analyses. It takes a simulation and ODE analysis as parameters (which are required to have the
same stop time and step size) and calculates the difference between their resulting data
points. All the enclosed commands then use this difference in the same way as
data points for the other analyses. For example, a comparison of the above
analyses:
\begin{code}
comparison(
    odes(stopTime=3.0,stepSize=0.001,density=1000){...},
    simulation(stopTime=3.0,stepSize=0.001,replications=1000){...}){
        plot(E[Processors:Processor0],E[Resources:Resource0]); 
}
\end{code}

\subsection{Commands and Functionality}

\GPEPA\ can plot results from these analyses itself or output raw data
by means of an optional file redirection.


The \texttt{plot} command provides direct plots of different arithmetic expressions
involving (higher order and joint) moments of component counts and
numerical constants. Each group--component pair is identified by the
syntax
\textit{group}\texttt{:}\textit{component}. A
moment is an expectation, with syntax \texttt{E[]}, of products of
these, for example \texttt{E[G1:C1\^{}2 G2:C2]}. Several shorthands
are provided, such as the variance represented by \texttt{Var[G:C]}
(which stands for \texttt{E[G:C\^{}2]-E[G:C]\^{}2}), and
\texttt{Cov[G1:C1,G2:C2]} which represents the covariance of two
component count variables and the $n$th central moment represented by
\texttt{Central[G:C,n]}.
The following commands are examples of the above:
\begin{code}
plot(Var[Processors:Processor1],Var[Resources:Resource1]);
plot(Central[Processors:Processor1,4]);
plot(E[Resources:Resource0]+E[Resources:Resource1]); 
\end{code}
The \texttt{plotSwitchpoints} command inside an ODE analysis
visualises distance from switch points in the ODE approximation. It
obtains a set of all the occurrences of the $\min$ function (containing only
moments of order upto a given integer argument) in the
moment ODEs. For each data point from the ODE analysis, the command
plots the difference between the two arguments of each of the $\min$
occurrences. The switch points then correspond to the zero points of
this plot. 

One of the reasons the \texttt{plot} command provides arithmetic expressions of
the moments is to give \GPEPA\ the flexibility to obtain approximations of passage
times. As described in~\cite{pepa-odes-passage}, the passage time approximations
and the corresponding bounds can be expressed as certain functions of the (higher
order) moments. We will
explore this functionality in a later paper, but suffice to say that
the accuracy of the passage-time approximations depends critically on
the ODE approximation of first and higher moments.






Some example plots showing error comparisons in variances for the
processor/resource model can be found in
Appendix~\ref{sec:procres_gpepa}. For further details on \GPEPA, including 
random model generation for testing purposes, we refer
the interested reader to the freely available source code obtainable
from~\cite{GPA}.

\section{Numerical Investigation: Two-stage Client/Server}
\label{sec:investigations}
In this section, we investigate empirically the nature of the ODE
approximations as mentioned in \sectref{sec:approx} and the
convergence results from \sectref{sec:theoretical}. More specifically,
we will check that the simulation variances converge to the ODE
approximations as the initial component populations scale up.  We also
look at higher moments to suggest a possible result analogous to the
one mentioned in \sectref{sec:conv_var}. We will look at the error of
the ODE approximation of the moments and quantitative properties of
the convergence, in the context of the \emph{switch points} defined in
\sectref{sec:switch}.

In order to investigate this, we use two models - one which does not
stay near a switch point in any large time interval (we will
informally refer to it as to the \emph{occasionally switching} model)
and one which steadily stays near a switch point for a longer period
of time (we will informally refer to it as to the \emph{persistently
  switching} model).

\begin{figure}[hb]
\begin{center}
  \subfloat[Switch point plot for Model
  A]{\includegraphics[scale=0.60]{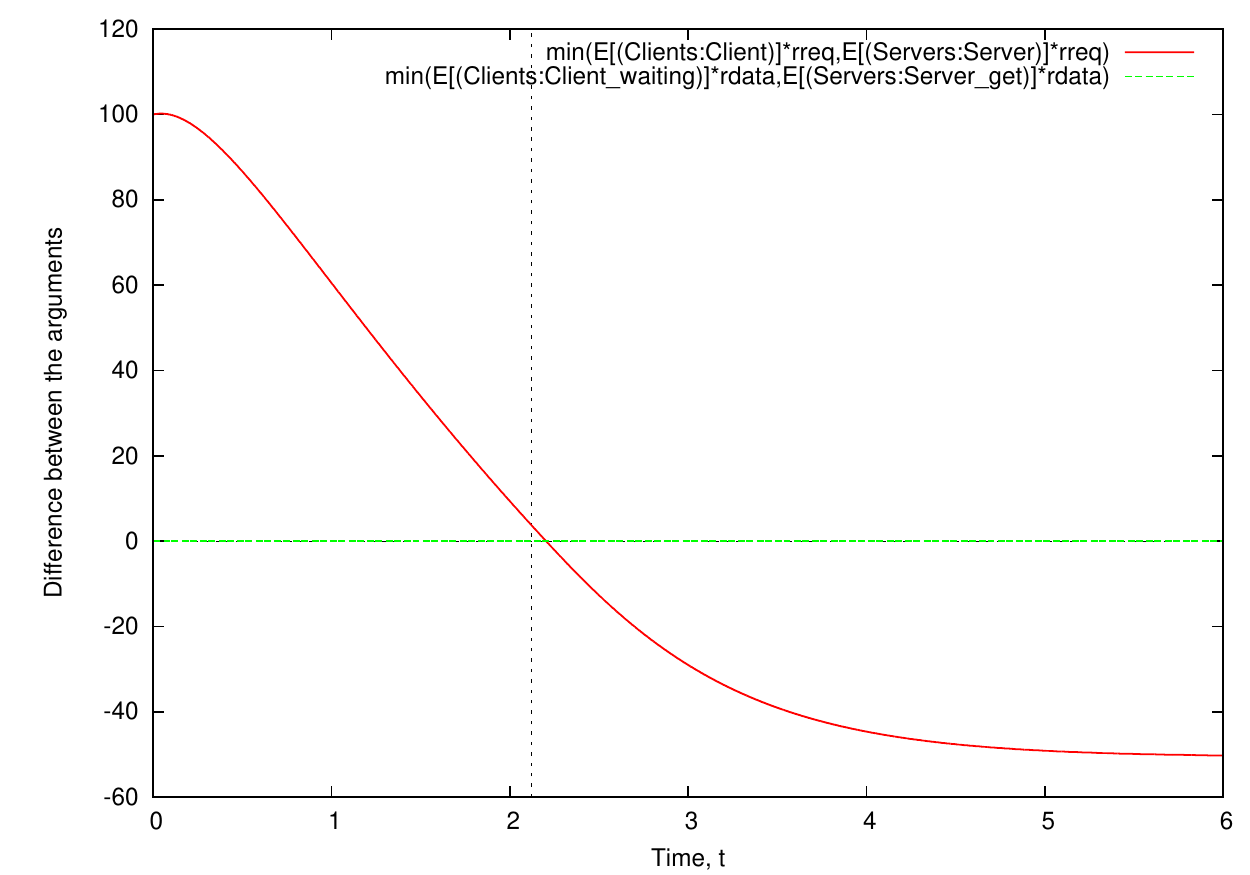}}
  \subfloat[Switch point plot for Model
  B]{\includegraphics[scale=0.60]{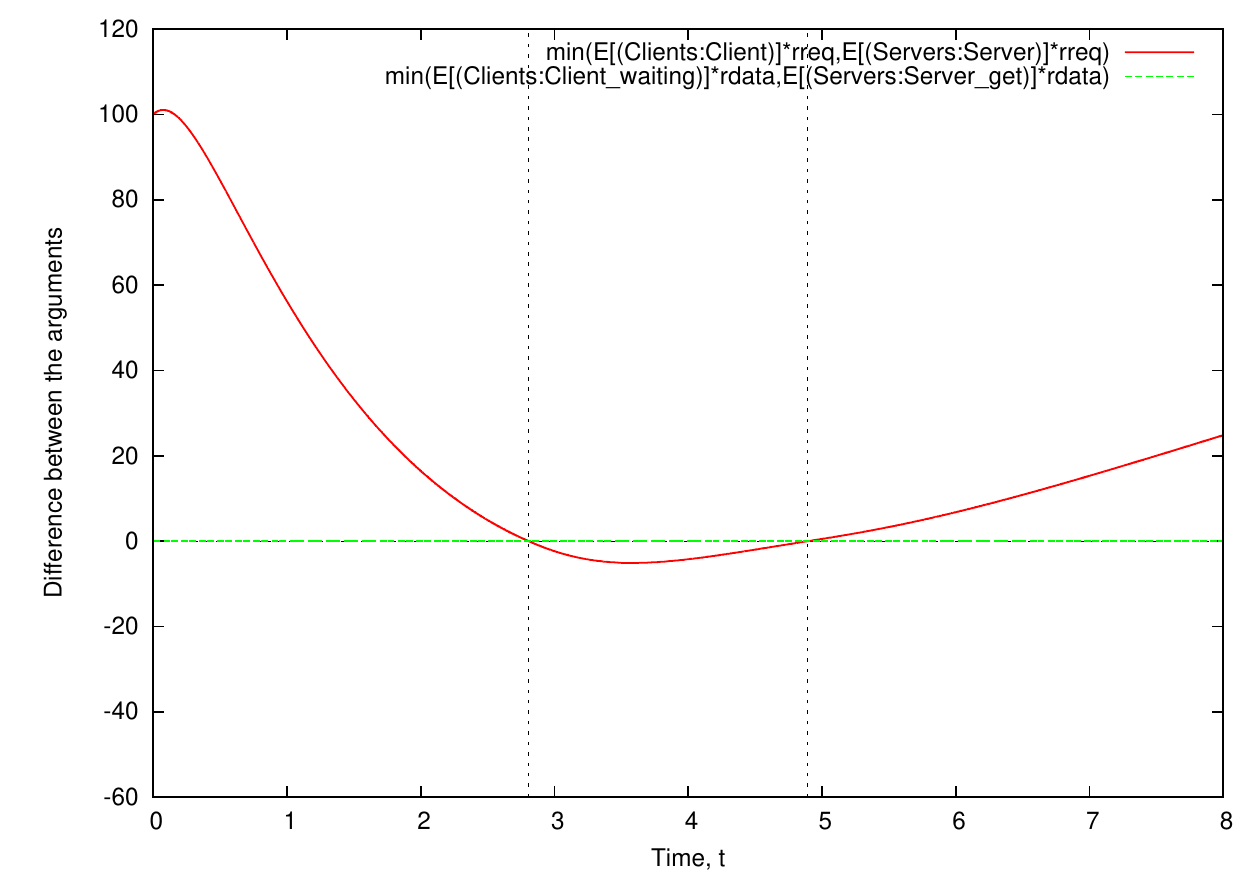}}
\end{center}
\caption{First order switch points for the two-stage client/server models}
\label{fig:twostage_switch}
\end{figure}

In both cases we will use the two-stage client/server model from
\sectref{sec:clientserver}, with two sets of parameters and taking the
client population, $c=100$, and the server population, $s=50$:
\begin{align*}
\text{Model A}:&\qquad r_\mathit{req} = 2.0&\qquad r_\mathit{break} = 0.1&\qquad
r_\mathit{think}=0.20&\qquad 
r_\mathit{data}=1.0&\qquad r_\mathit{reset} = 2.0\\
\text{Model B}:&\qquad r_\mathit{req} = 2.0&\qquad r_\mathit{break} = 0.3&\qquad
r_\mathit{think}=0.35&\qquad 
r_\mathit{data}=2.0&\qquad r_\mathit{reset} = 0.05
\end{align*}
The switch point behaviour for these examples can be seen on
\figref{fig:twostage_switch}.  There are two possible sources of
switch points in the client/server model, each corresponding to an
instance of cooperation in the model. One, the term
$\min(v_\pepa{Client}(t)\cdot r_\mathit{req},v_\pepa{Server}(t)\cdot
r_\mathit{req})$, comes from the cooperation when a client establishes
connection with a server. Another,
$\min(v_{\pepa{Client}_\pepa{waiting}}(t)\cdot
r_\mathit{data},v_{\pepa{Server}_\pepa{get}}(t)\cdot
t_\mathit{data})$, comes from the cooperation when a client retrieves
data from an available server.

For the $\min$ term involving $\pepa{Server}_\pepa{get}$ and
$\pepa{Client}_\pepa{waiting}$, the model hits infinitely many switch points.
These do not influence the error of the approximation since the two
corresponding counting processes are stochastically identical, so there is no error in
the corresponding expectation approximation.  For the $\min$ term involving the
$\pepa{Client}$ and $\pepa{Server}$ components, in the first, simpler case, the
model hits one switch point at time $t\simeq 2.1$, but does not stay around any
switch point for a period of time. The second model hits two switch points when
$t\simeq2.8$ and $t\simeq4.8$ and stays close to a switch point in the time
interval between them. 


In the following sections, we will look at the errors in moment
approximation for both of the above models and compare them in the
context of these switch points. We will look at the expectation,
variance, skewness and kurtosis (the standardised third and fourth
central moments) of the component counts for each model and its
respective versions with initial populations scaled by a factor of
$n=1,4,16$ and $64$. We investigate how the scale influences the error
of the ODE approximation. For each scale, we plot the difference
between the moments from simulation and their approximation, specially
near the times where a switch point is encountered. It is worth noting
that these times are the same across all $n$, since the ODEs for the
means are scale invariant, as was mentioned in
\sectref{sec:theoretical}.

In line with the considerations from \sectref{sec:theoretical}, we normalise
the error of expectation and variance approximations dividing by the total
component population. We plot the errors of the skewness and kurtosis
approximations (standardised third and fourth central moments, respectively)
without any additional normalisation. 

\subsection{Model A: Occasionally switching model} 

\begin{figure}[hbt]
\begin{center}
\subfloat[ODE approximation: mean]{\includegraphics[scale=0.50]{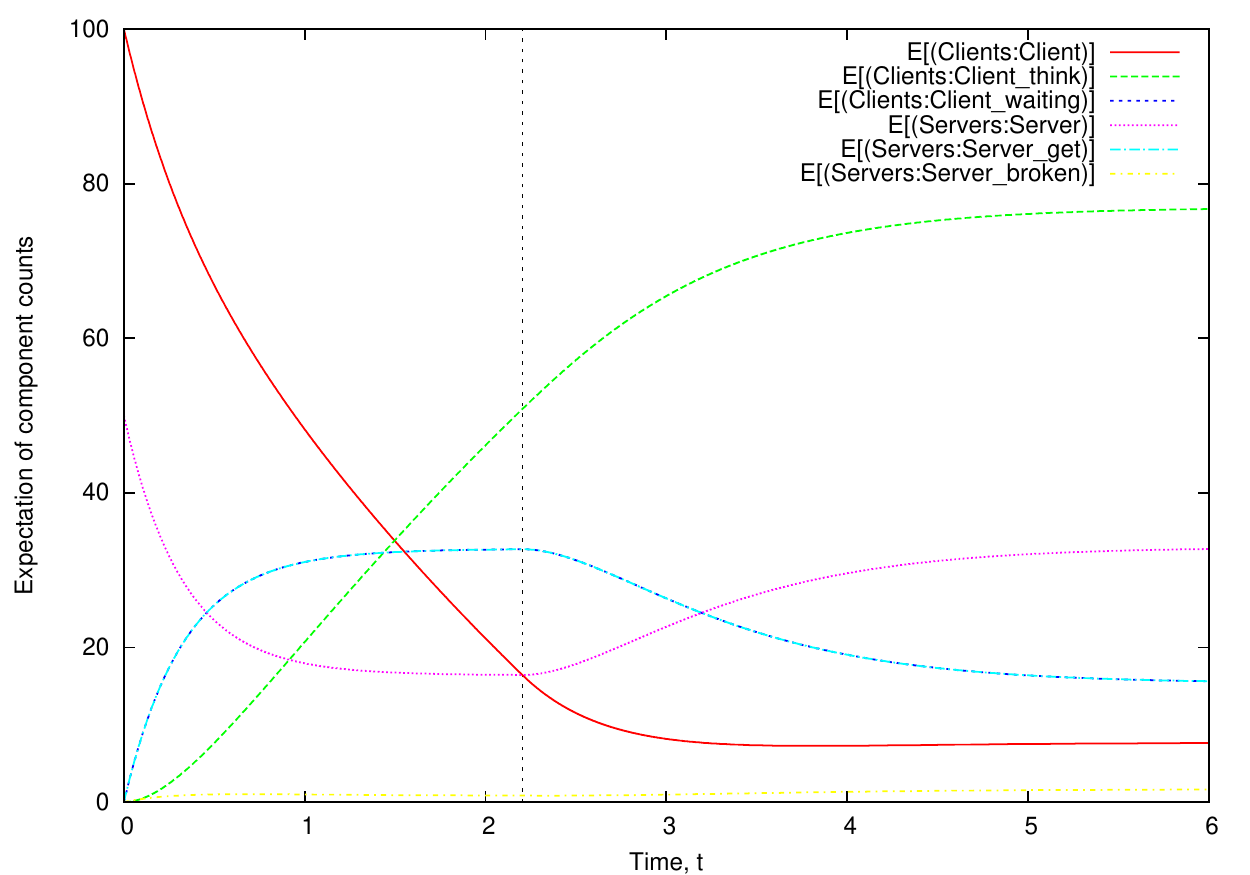}}
\subfloat[Simulation: mean]{\includegraphics[scale=0.50]{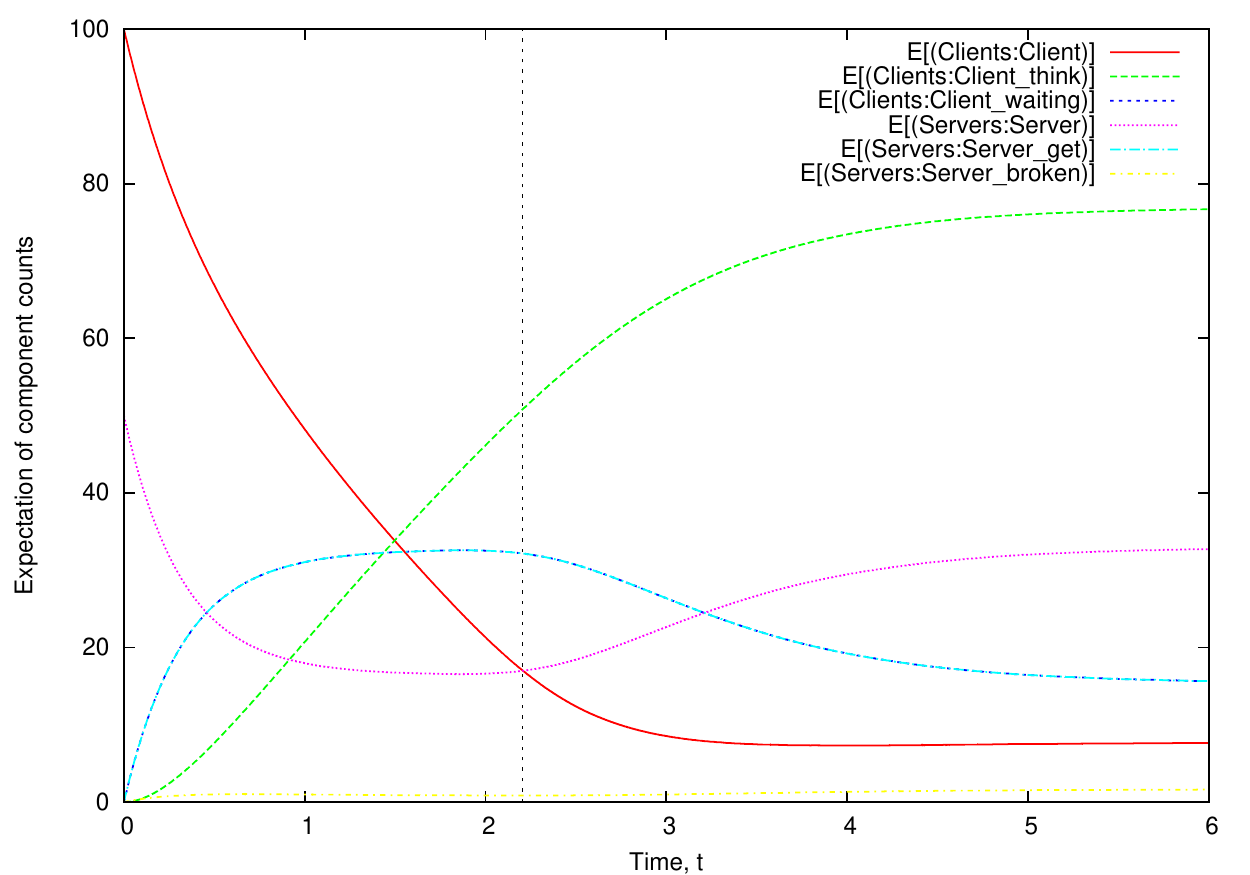}}\\
\subfloat[ODE approximation:  variance]{\includegraphics[scale=0.50]{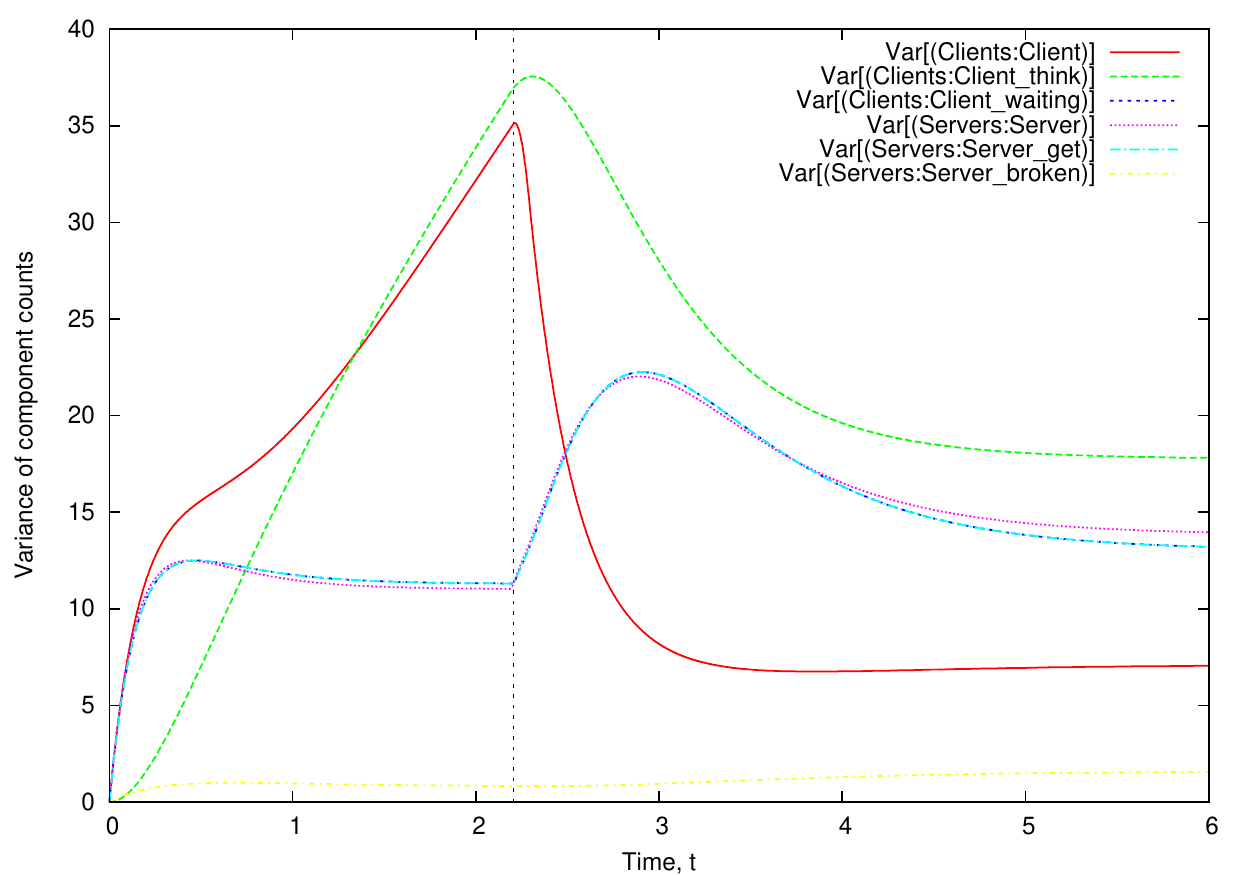}}
\subfloat[Simulation: variance]{\includegraphics[scale=0.50]{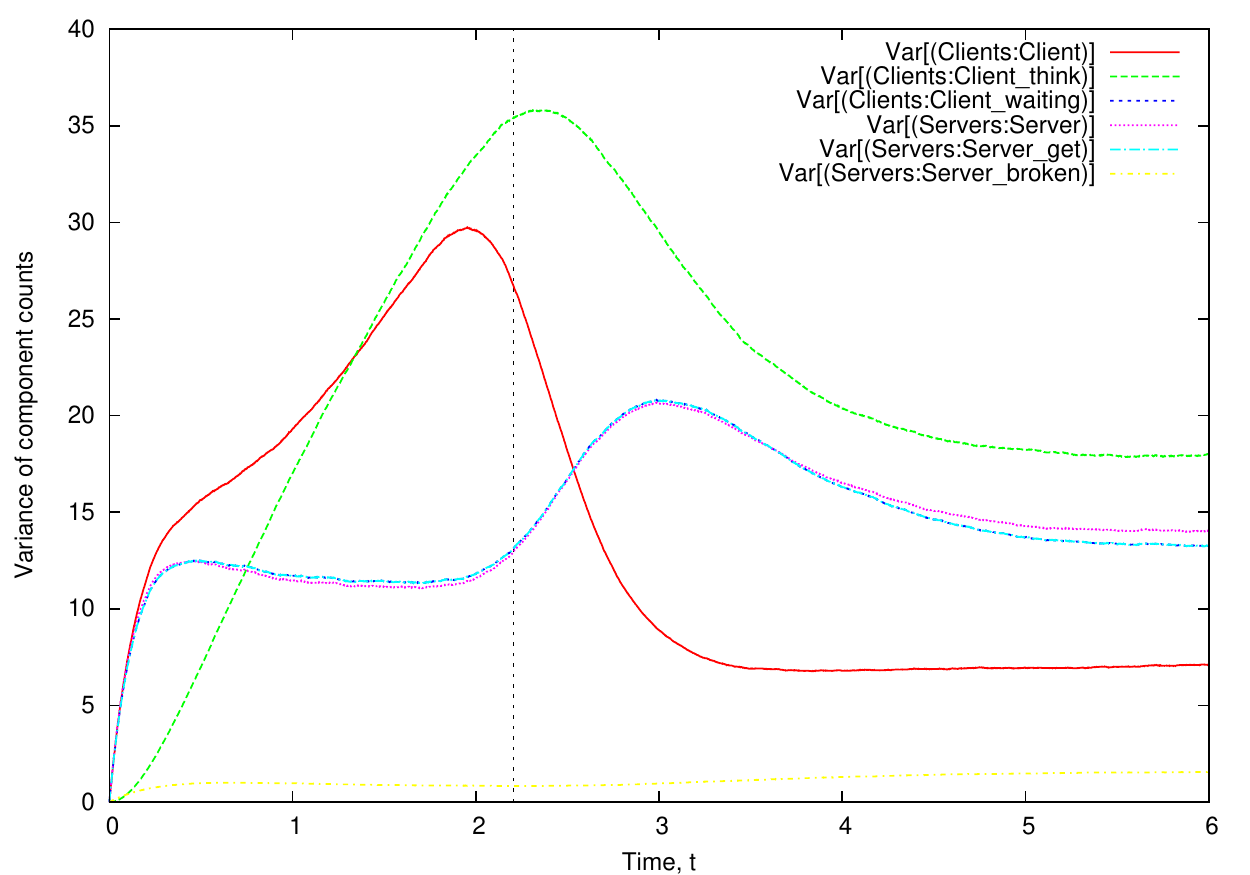}}\\
\subfloat[ODE approximation: skewness]{\includegraphics[scale=0.50]{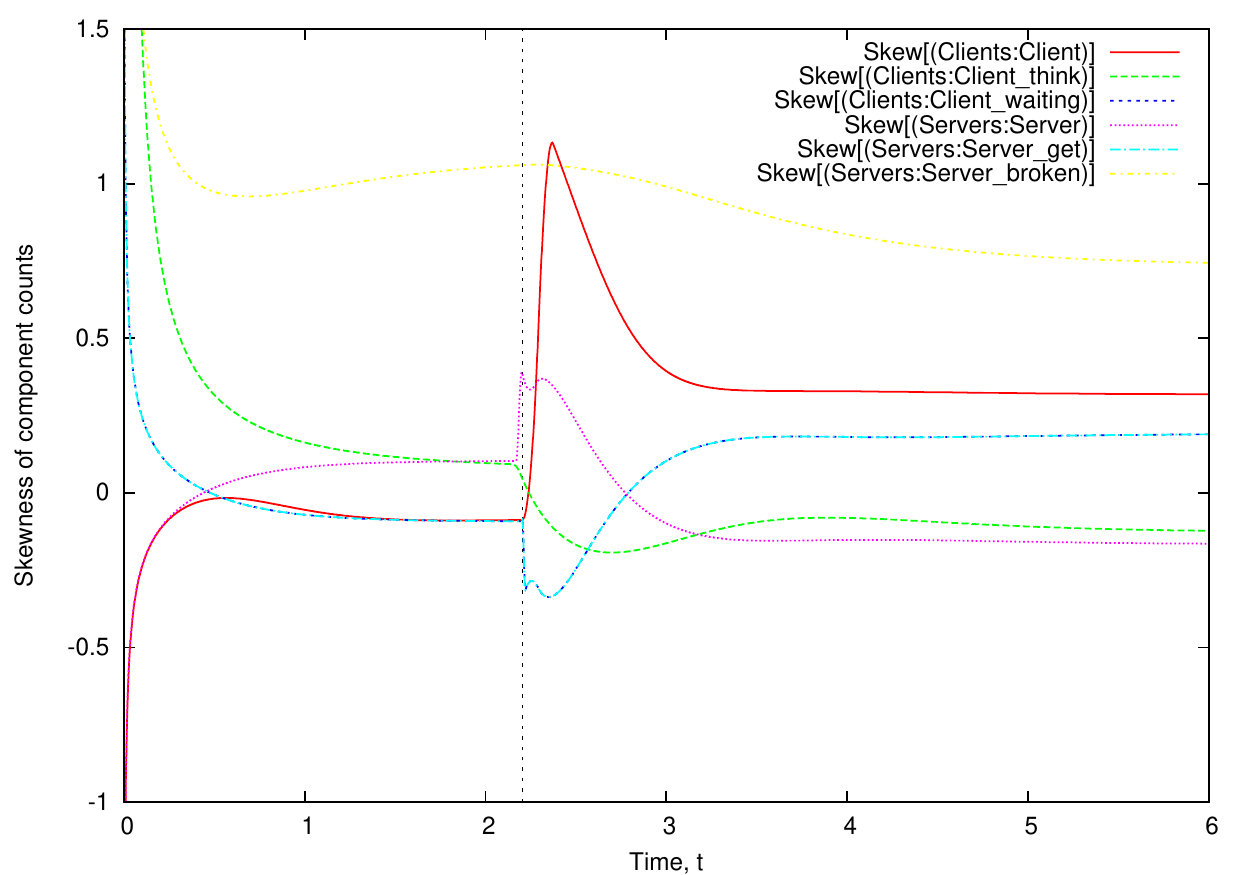}}
\subfloat[Simulation: skewness]{\includegraphics[scale=0.50]{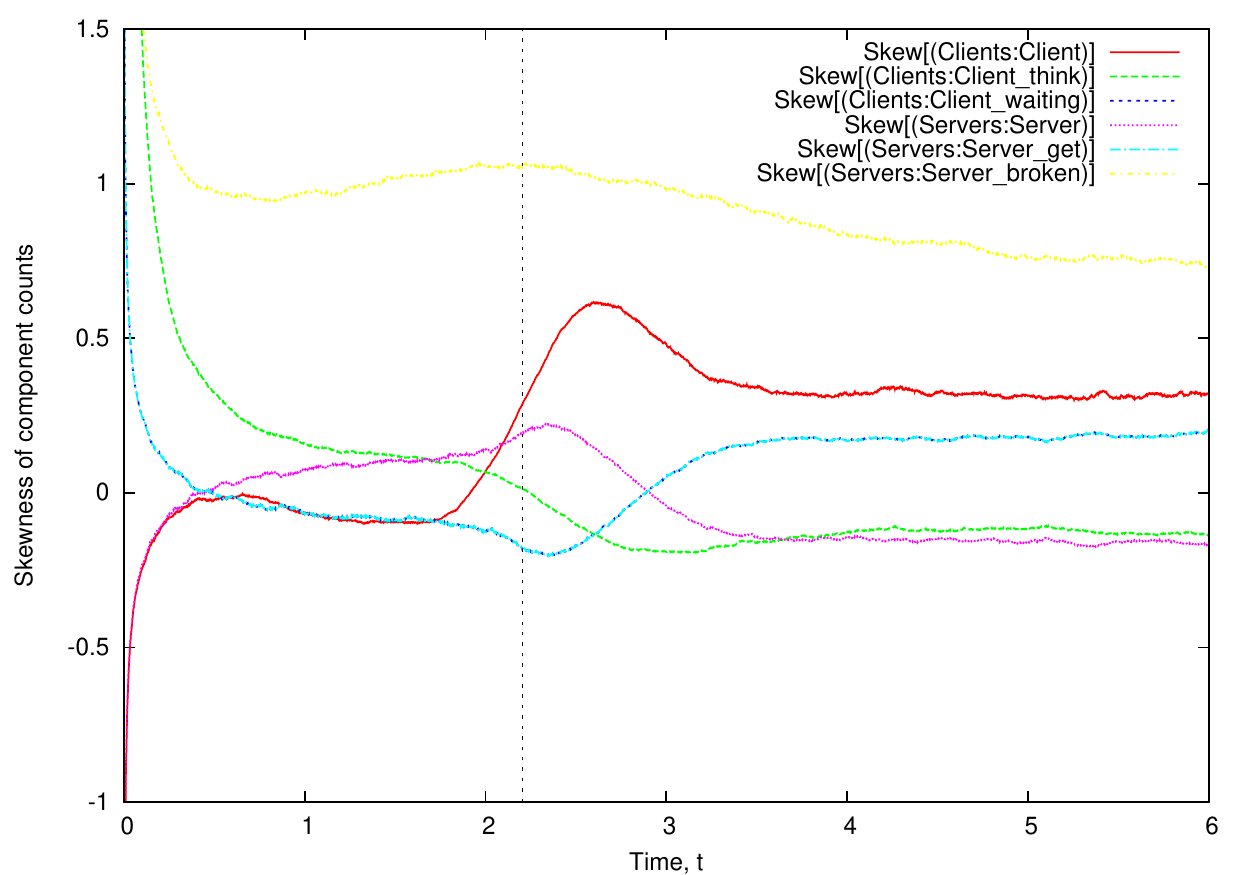}}\\
\end{center}
\caption{Moments from simulation and approximation for the two-stage
  client/server model, Model A.}
\label{fig:twostage_moments}
\end{figure}

First, we look at the simpler case, Model A, in which the
client/server model does not stay near a switch point for a longer
period of time. We start by comparing side by side, for the first
three central moments, the simulation average with its ODE
approximation in \figref{fig:twostage_moments}.

For kurtosis (not shown), similar to the skewness, the approximation is quite accurate when the
model is not near a switch point, but errors are visible otherwise. 

\begin{figure}[hbt]
\begin{center}
\subfloat[Normalised error: mean]{\includegraphics[scale=0.50]{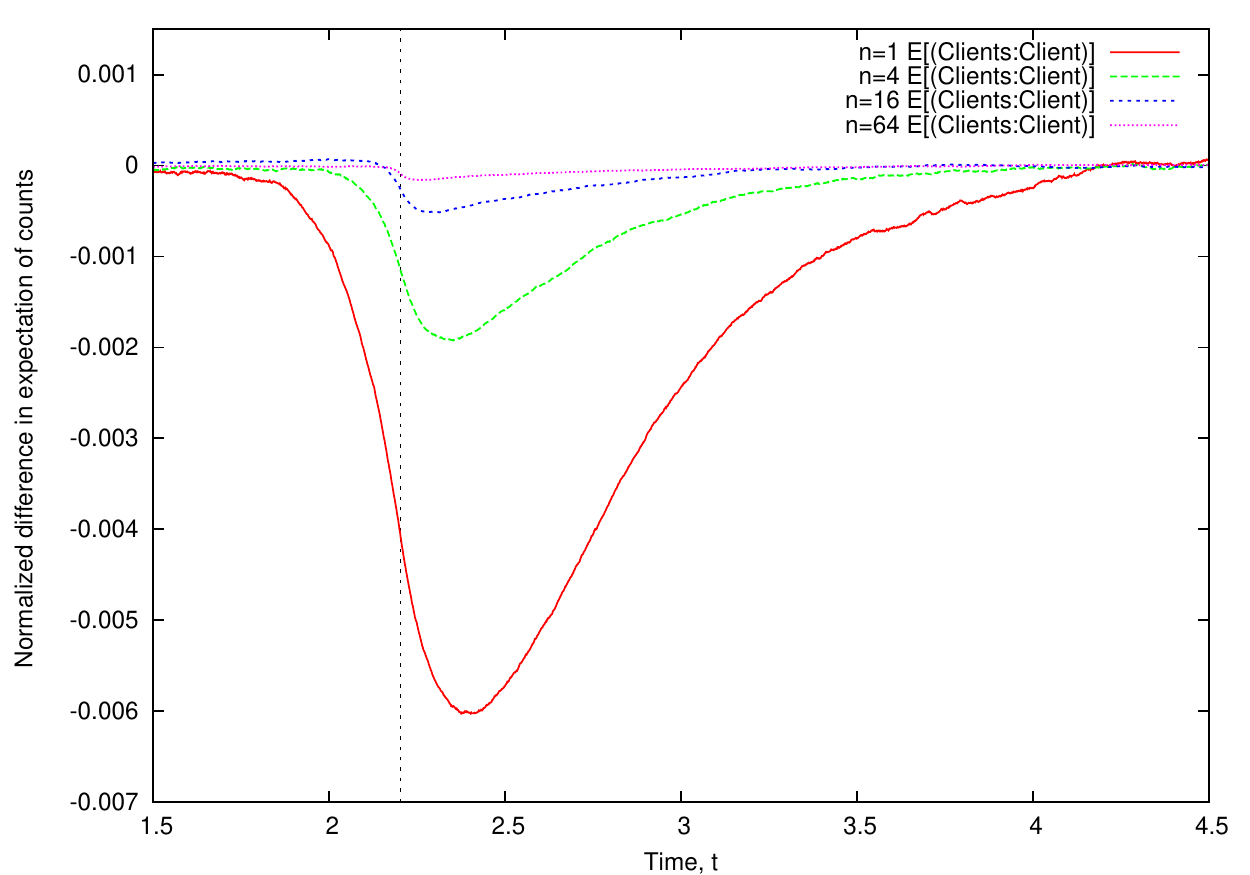}}
\subfloat[Normalised error: variance]{\includegraphics[scale=0.50]{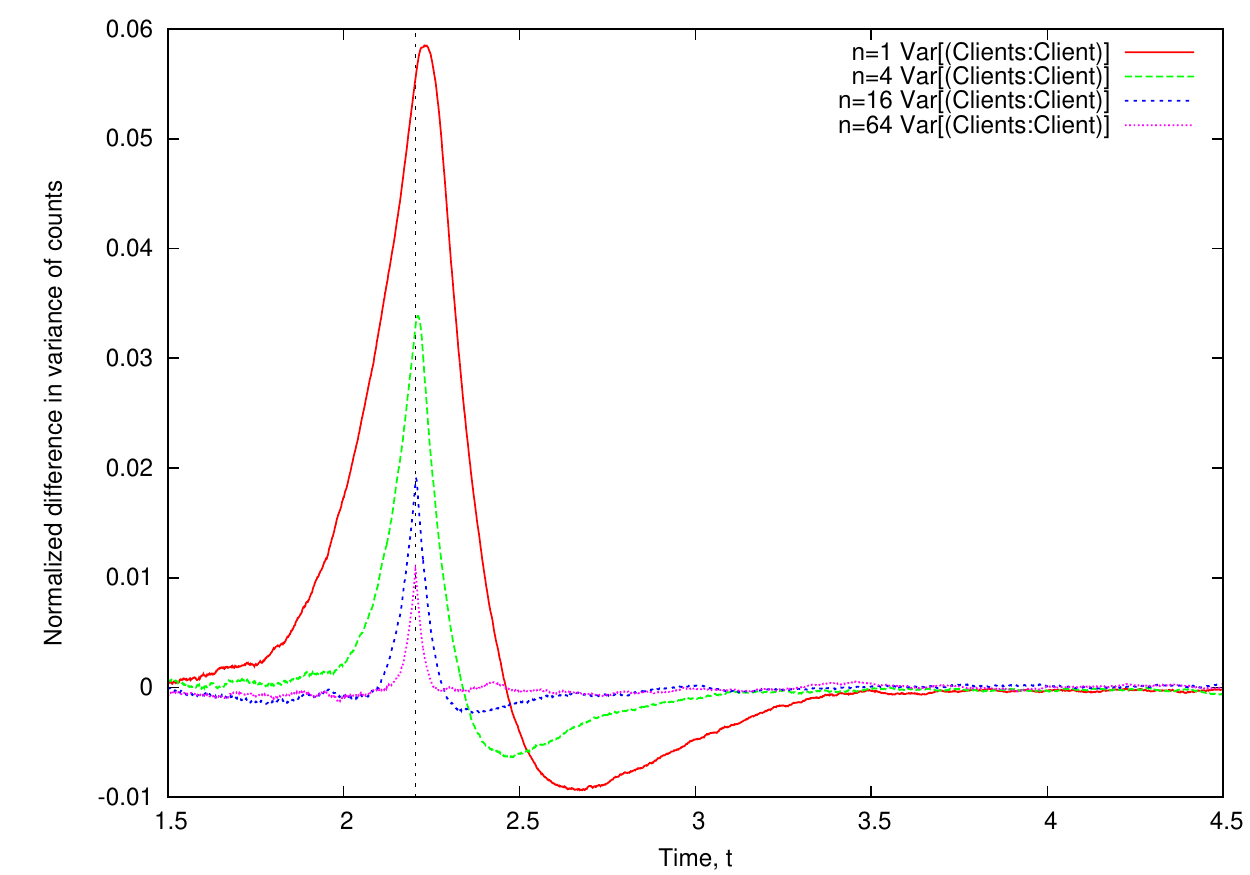}}\\
\subfloat[Absolute error: skewness]{\includegraphics[scale=0.50]{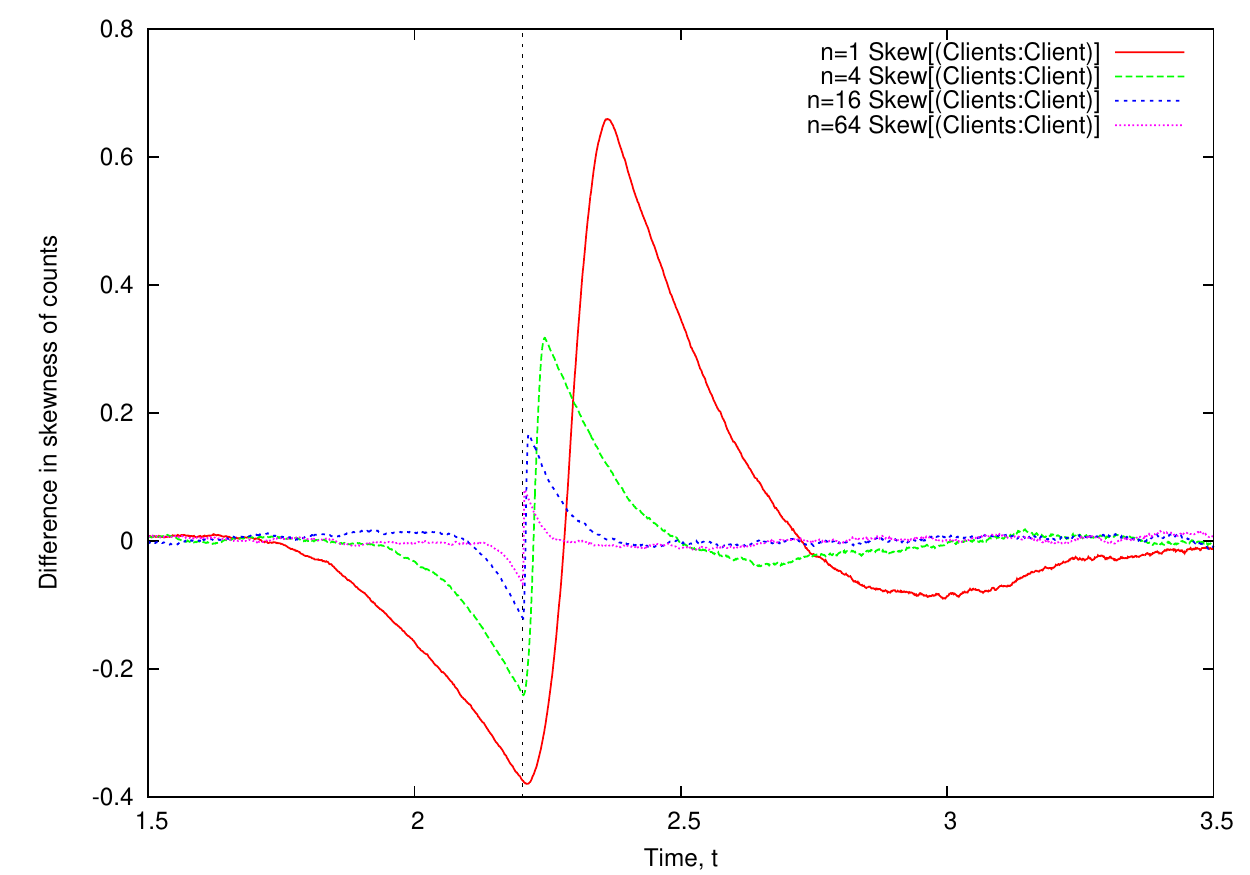}}
\subfloat[Absolute error: kurtosis]{\includegraphics[scale=0.50]{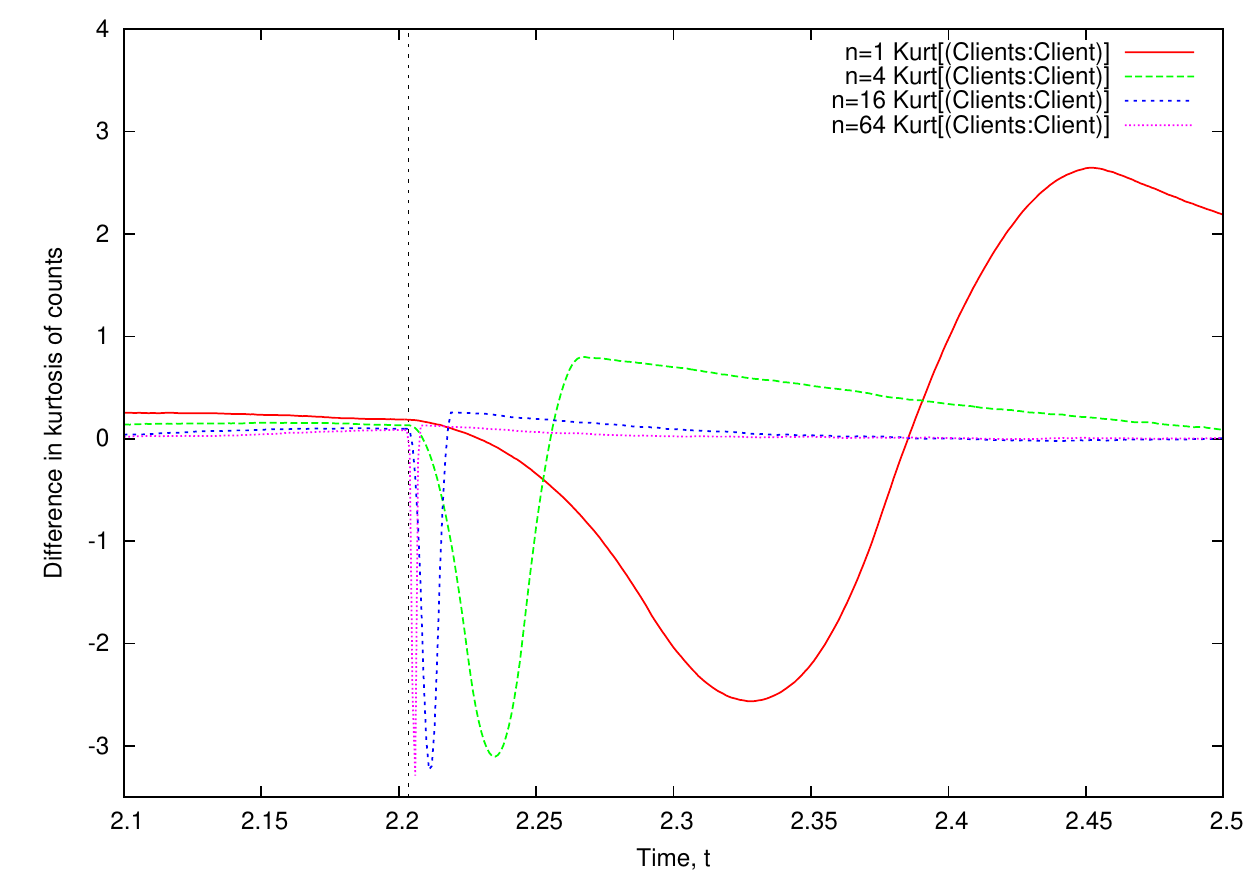}}\\
\end{center}
\caption{The influence of scaling of Model A on the normalised error around the switch point events}
\label{fig:twostage_error}
\end{figure}

For all four moments the ODE approximation is qualitatively close to
the simulation average. However, there are visible quantitative
differences starting at the variance and getting worse with the
skewness and kurtosis. This is especially noticeable for the moments
involving the number of $\pepa{Client}$ components and in the time
interval around the first order switch point from
\figref{fig:twostage_switch} (shown by the vertical line on the
plots in \figref{fig:twostage_moments}). We plot the actual error of the approximation (using the
\GPEPA\ comparison analysis) for these moments, for each of the scaled
versions of the model, in \figref{fig:twostage_error}.
 
\subsection{Model B: Persistently switching model}

\begin{figure}[hbt]
\begin{center}
\subfloat[ODE approximation: mean]{\includegraphics[scale=0.50]{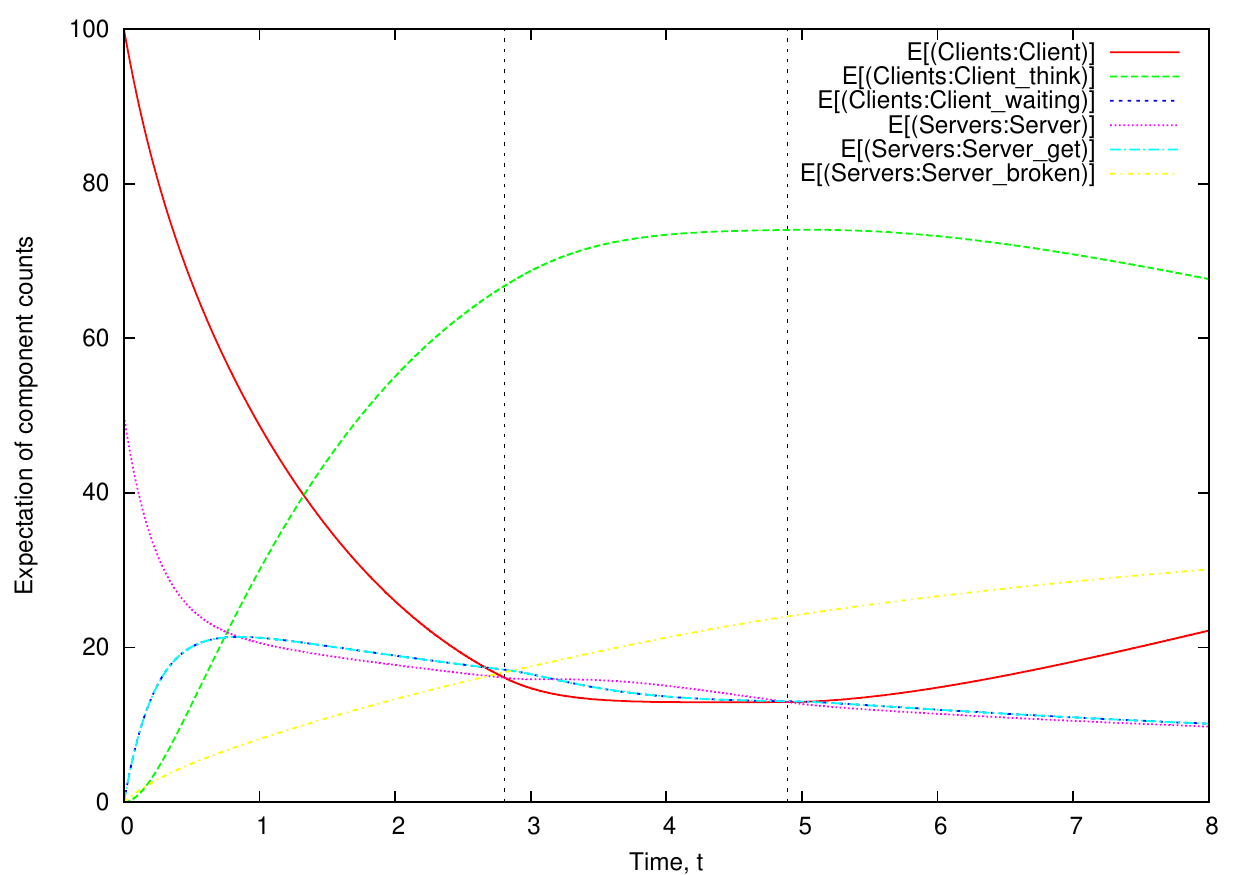}}
\subfloat[Simulation: mean]{\includegraphics[scale=0.50]{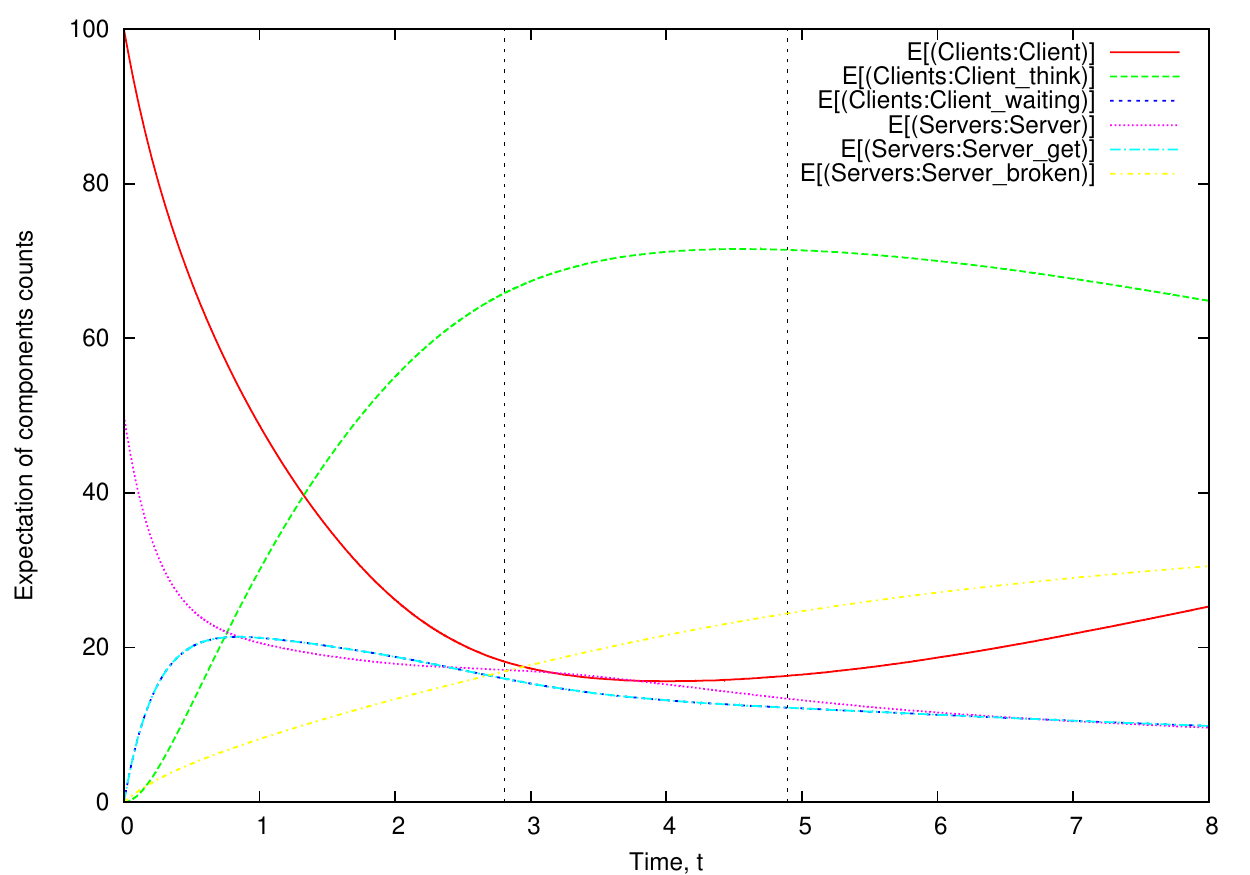}}\\
\subfloat[ODE approximation: variance]{\includegraphics[scale=0.50]{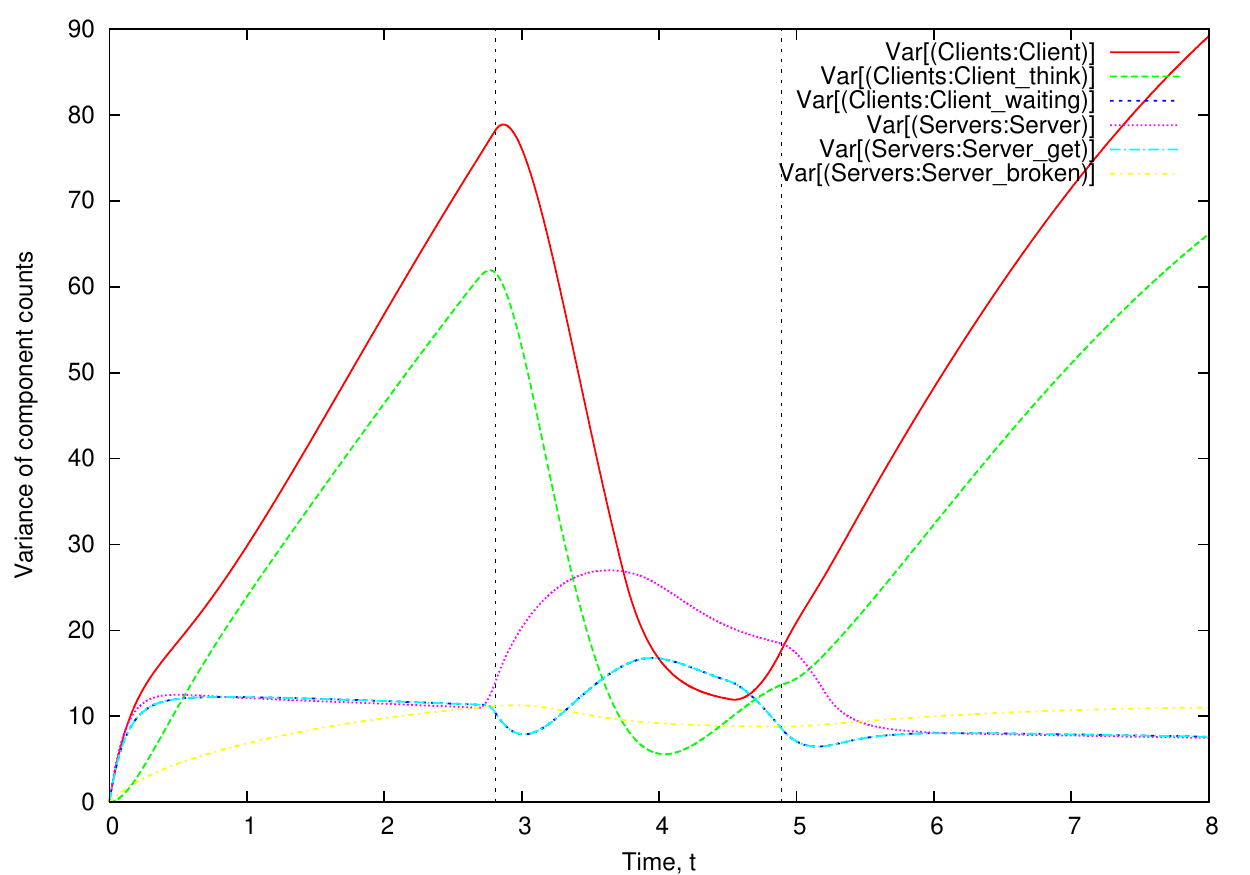}}
\subfloat[Simulation: variance]{\includegraphics[scale=0.50]{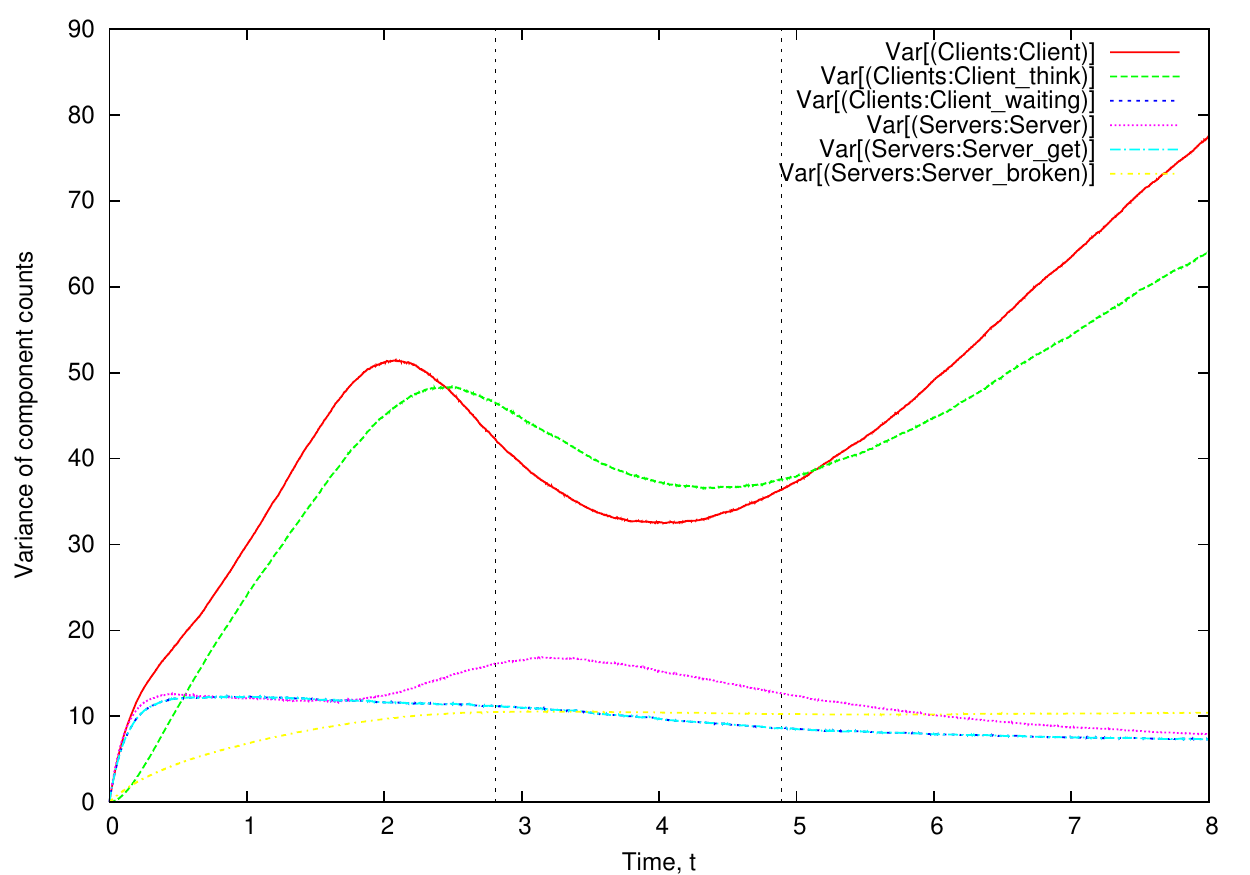}}\\
\end{center}
\caption{Moments from simulation and approximation for the two-stage
  client/server model, Model B. The ODE approximation is quite
  accurate for the mean, but large differences are visible for the
  variance.}
\label{fig:twostageWorse_moments}
\end{figure}

\Figref{fig:twostageWorse_moments} shows the mean and variance for the
case where the model stays near a switch point for a longer interval
of time in Model B. The mean seems quite accurate, but the error in variance
approximation is high around the interval where the model stays near a
switch point.

\begin{figure}[hbt]
\begin{center}
\subfloat[Normalised error: mean]{\includegraphics[scale=0.50]{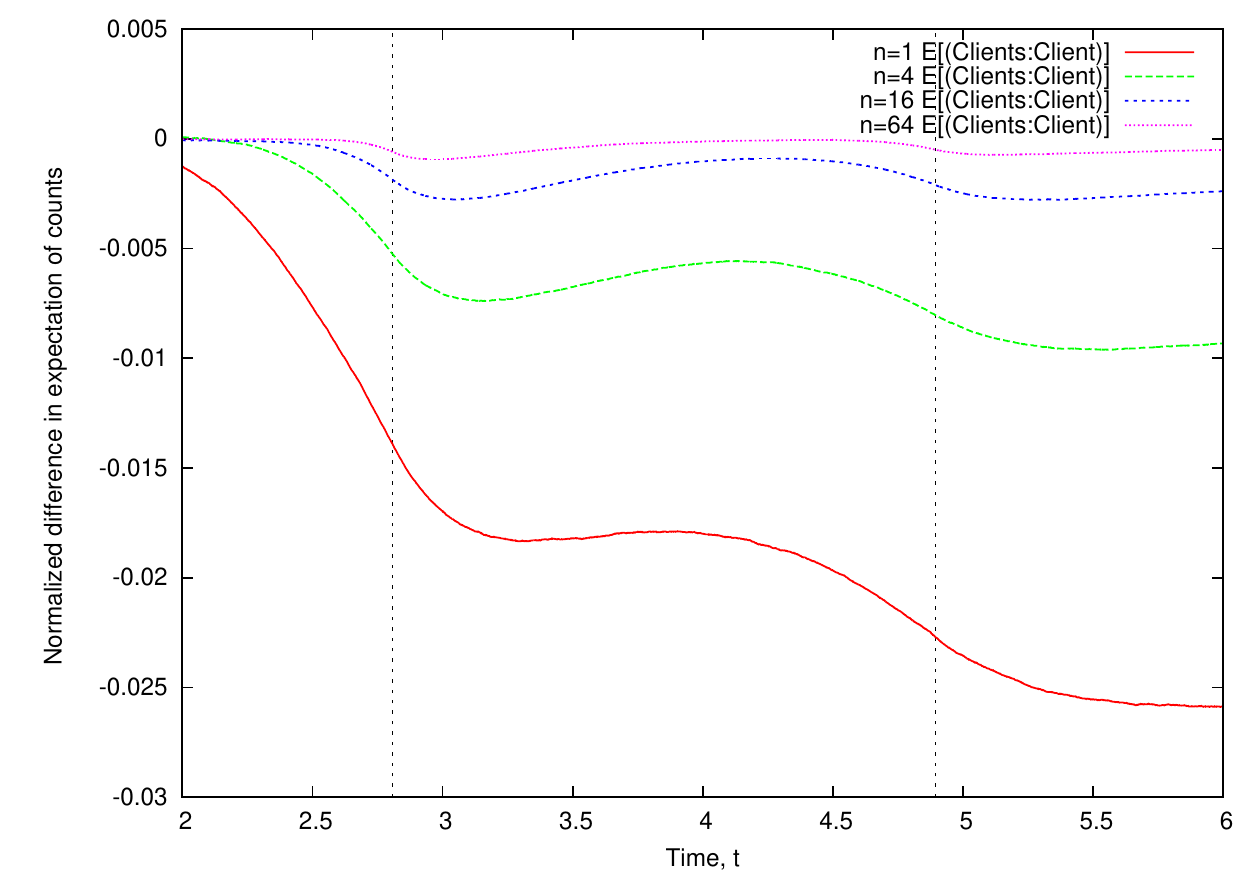}}
\subfloat[Normalised error: variance]{\includegraphics[scale=0.50]{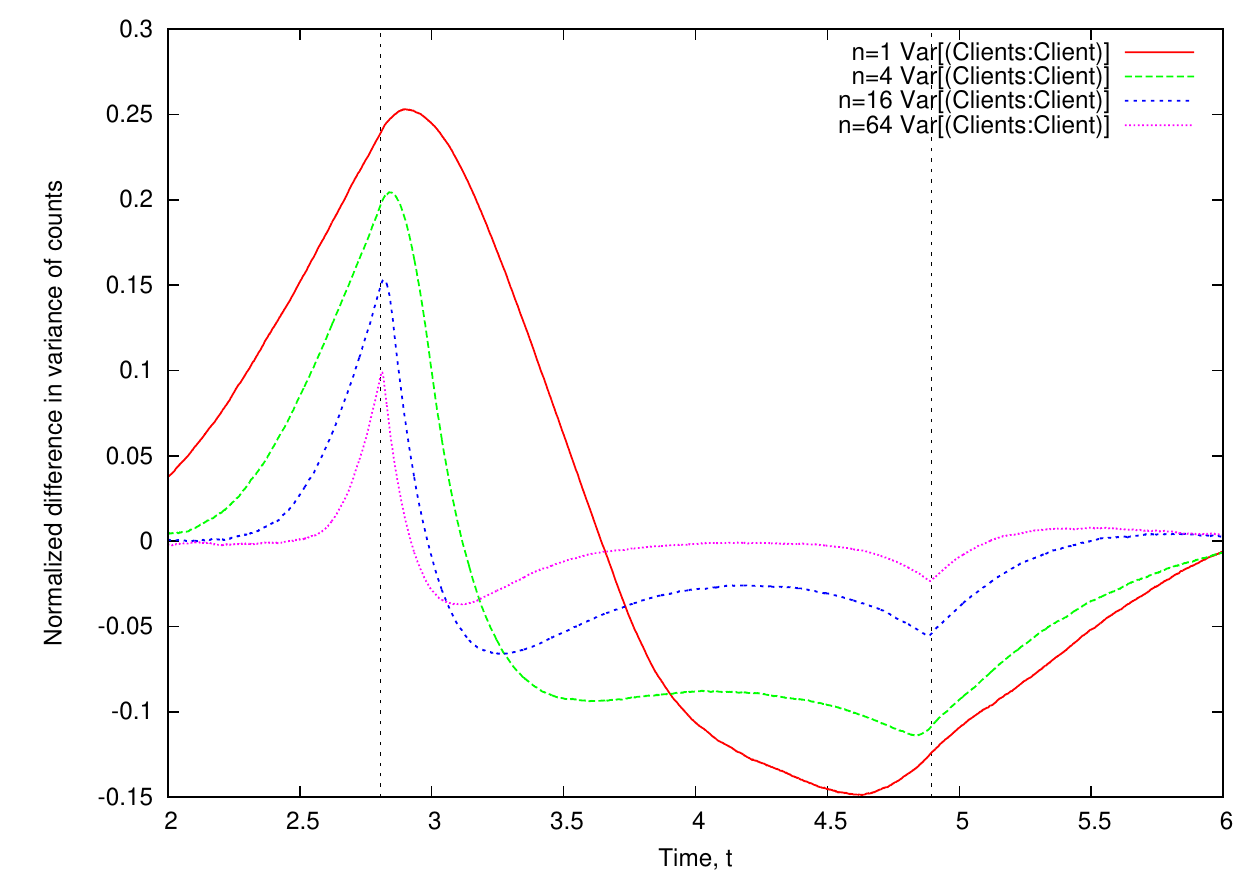}}\\
\subfloat[Absolute error: skewness]{\includegraphics[scale=0.50]{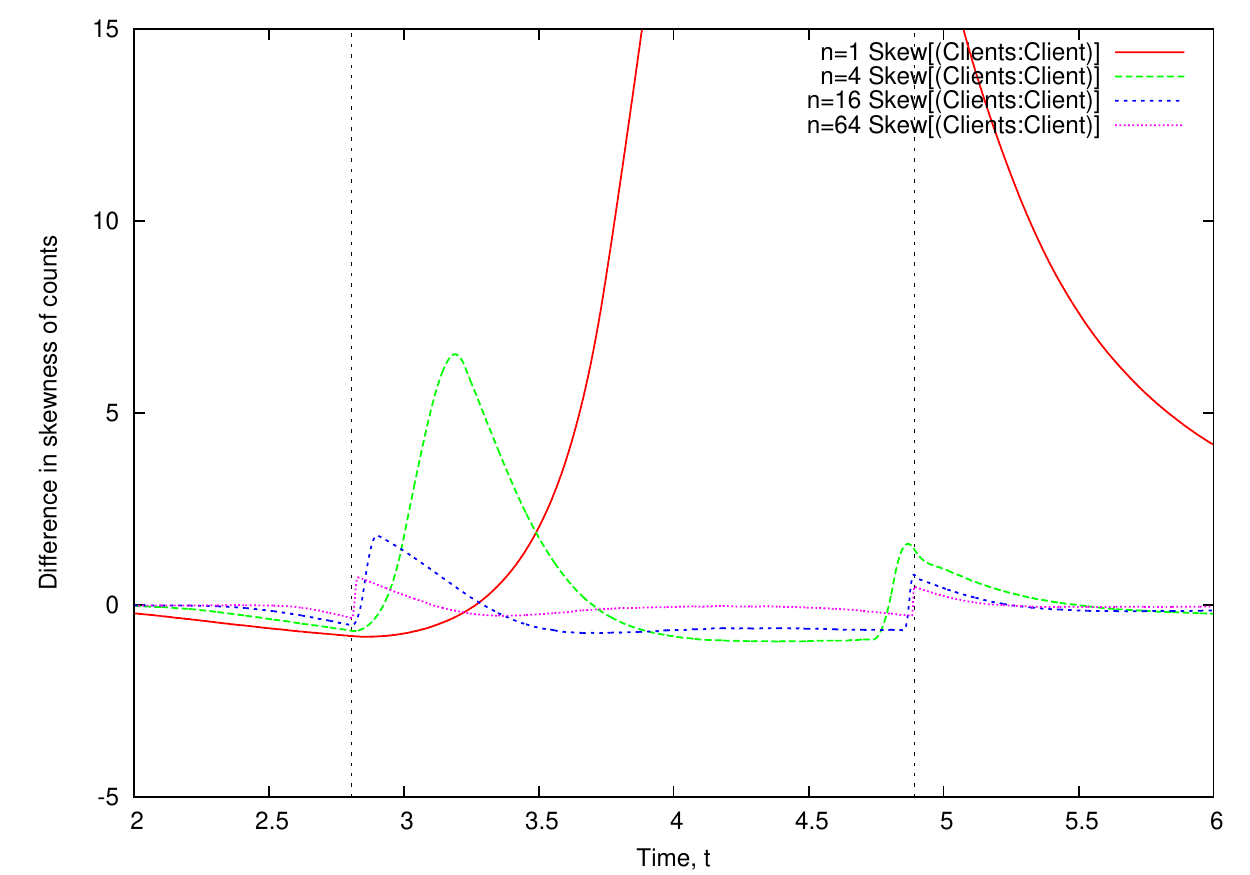}}
\subfloat[Absolute error: kurtosis]{\includegraphics[scale=0.50]{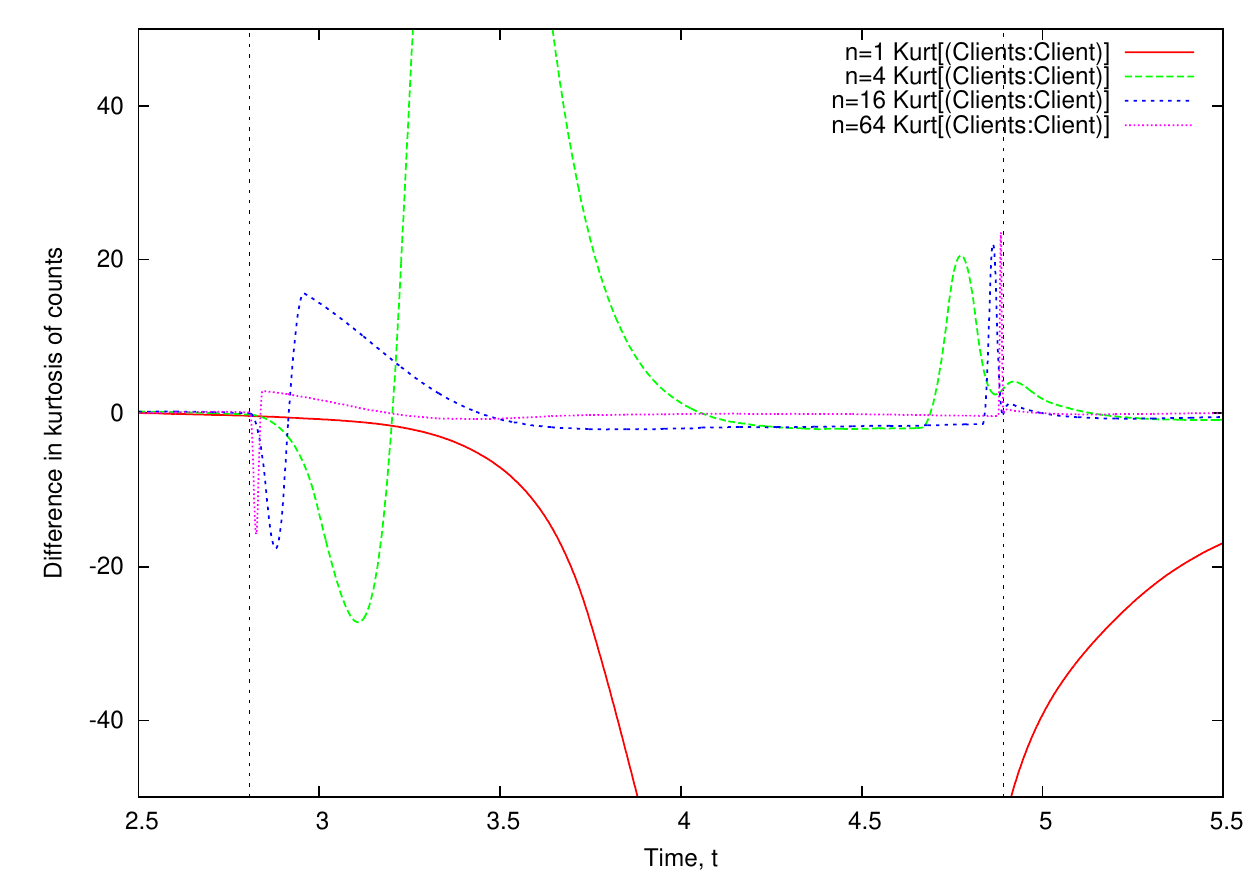}}\\
\end{center}
\caption{The influence of scaling of Model B on the normalised error around the switch point events}
\label{fig:twostageWorse_error}
\end{figure}

\Figref{fig:twostageWorse_error} looks at the error more closely and plots the
difference between the moments from simulation and their ODE approximation for
the $\Client$ component, for different scales of Model B.
It can be seen that the normalised error in
\figref{fig:twostageWorse_error} is much higher than in the case of
Model~A. We believe this is caused by the fact that the model is
closer to the switch points.  However, in both cases we can confirm
that the error for the mean and variance seems to be going to zero in
the scale limit, but concentrated most around a switch point.
The same seems to be the case for the skewness. For kurtosis, it seems that the
error does not necessarily get smaller in value, but the interval where it
appears seems to get smaller with increased scale. 

\Sectref{sec:theoretical} shows that the marginal
distributions of component counts become normal as the populations are
increased. The normal distribution has a constant skewness and
kurtosis (0 and 3 respectively). We can check the plots for the
simulations to verify this. This seems to be the case for the ODE
approximation, except perhaps for the points concentrated around the
switch point times.

\section{Conclusion and future work}

We introduced a tool, \GPEPA~\cite{GPA}, that for the first time makes
available the ODE approximations of higher moments to a wide range of
models described in an extension of the PEPA stochastic process
algebra. 
We used the tool to carry out investigations into the nature of the
ODE approximation, that will help the modellers to detect errors
without running computationally more expensive stochastic simulations.
We theoretically justified that the variance of the component counts
converges to the ODE approximation as the initial populations are
scaled and with the help of our tool verified this for a simple
example.

We observed, for a model where the resulting differential equations
are piecewise linear, that the error is influenced by how closely the
model stays near the switch points during its time evolution. If the
model only crosses switch points at certain points of time and does
not stay near any during the rest of the time, then the error is
concentrated tightly around those switch point events. We also saw
that, with increasing scale of initial component populations, the
error in the ODE solution becomes even more tightly concentrated
around the switch point. If the model stays near switch points for
longer periods of time, the resulting error is much more severe and
decreases more slowly with increased scale of the initial populations.

These observations help us to assess the validity of the ODE
approximation without actually running the simulations. For a given
model, we can use \GPEPA\ to visualise the switch point behaviour of
the model and use the intuition gained from our investigations to say
whether the approximation is accurate. 

Moreover, the presented results and observations are not just specific
to the PEPA stochastic process algebra. The $\min$ functions with the
concept of switch points appear in situations when there is a
competition for multiple resources, for example multi-server queues or
many-server semantics stochastic Petri nets. Therefore we believe that
the gained insight is relevant to a wider area within performance
modelling of computer systems.

In future, we plan to develop methods that would be able to quantify
the error (\eg in terms of bounds obtained from the distance from a
switch point), thus making \GPEPA\ able to warn the modeller of the
potential magnitude of any error.
We also plan to investigate the sensitivity of the switch point
behaviour to changes in the rates and initial population parameters to
allow deliberate generation of accurate ODE approximation.




\bibliography{paper,jb}

\appendix

\section{PEPA Summary}
\label{sec:PEPA}

We briefly introduce PEPA~\cite{Hil96}, a simple stochastic process
algebra with sufficient expressiveness to model a wide variety of
systems.  As in various other process algebras, systems in PEPA are
represented as compositions of \emph{components} which undertake
\emph{actions} thus evolving into further components. Each action has
a \emph{rate}, interpreted as a parameter of an exponential random
variable that governs the delay associated with the action. This means
that the stochastic behaviour of the model can be described as a CTMC.

The basic building blocks of PEPA are \emph{sequential components}, described by the syntax 
\begin{equation*}
S  ::= (\alpha, r).S ~ \mid ~ S + S ~ \mid ~ C_S 
\end{equation*} 
where $C_S$ stands for a constant that defines a sequential component. Intuitively, the component $(\alpha,r).S$ can undertake the action $\alpha$ with rate $r$ and evolve into the component $S$. The
component $P+Q$ has a choice to undertake actions that both $P$ and $Q$ can undertake. 

The sequential components can be composed in parallel to form \emph{model components}, described by the syntax  
\begin{equation*}
P  ::= P \sync{L} P ~ \mid ~ P / L ~ \mid ~ C
\end{equation*} 
where $C$ is a constant defining any PEPA component.  The component
$P\sync{L}Q$ enables cooperation between $P$ and $Q$ by only allowing
$P$ and $Q$ to evolve together when undertaking an action from the
cooperation set $L$. The cooperation influences the rate of the common
evolution. PEPA assumes \emph{bounded capacity}: a component cannot
be made to perform an activity faster by cooperation, so the resulting
rate is the minimum of the cooperating rates. This is discussed in
more detail in~\cite{Hil96}.  For actions not in $L$, $P$ and $Q$ can
evolve independently with no influence on the rates. If $L$ is the
empty set, we use the shorthand $P\parallel Q$ meaning that $P$ and
$Q$ are purely concurrent and don't synchronise. We also use the
shorthand $P[n]$ for $n$ purely concurrent copies of $P$.

The syntax $P/L$ stands for action hiding. For simplicity, we will not
consider this feature in the presented work, however, it should be
straightforward to include using the results from~\cite{pepa-odes-ck}.

The full details of the operational semantics of PEPA can be found in~\cite{Hil96}.

\section{Processor/Resource Variance Plots}
\label{sec:procres_gpepa}

\begin{figure}[hbt]
\begin{center}
\subfloat[Normalised difference in
variance]{\includegraphics[scale=0.60]{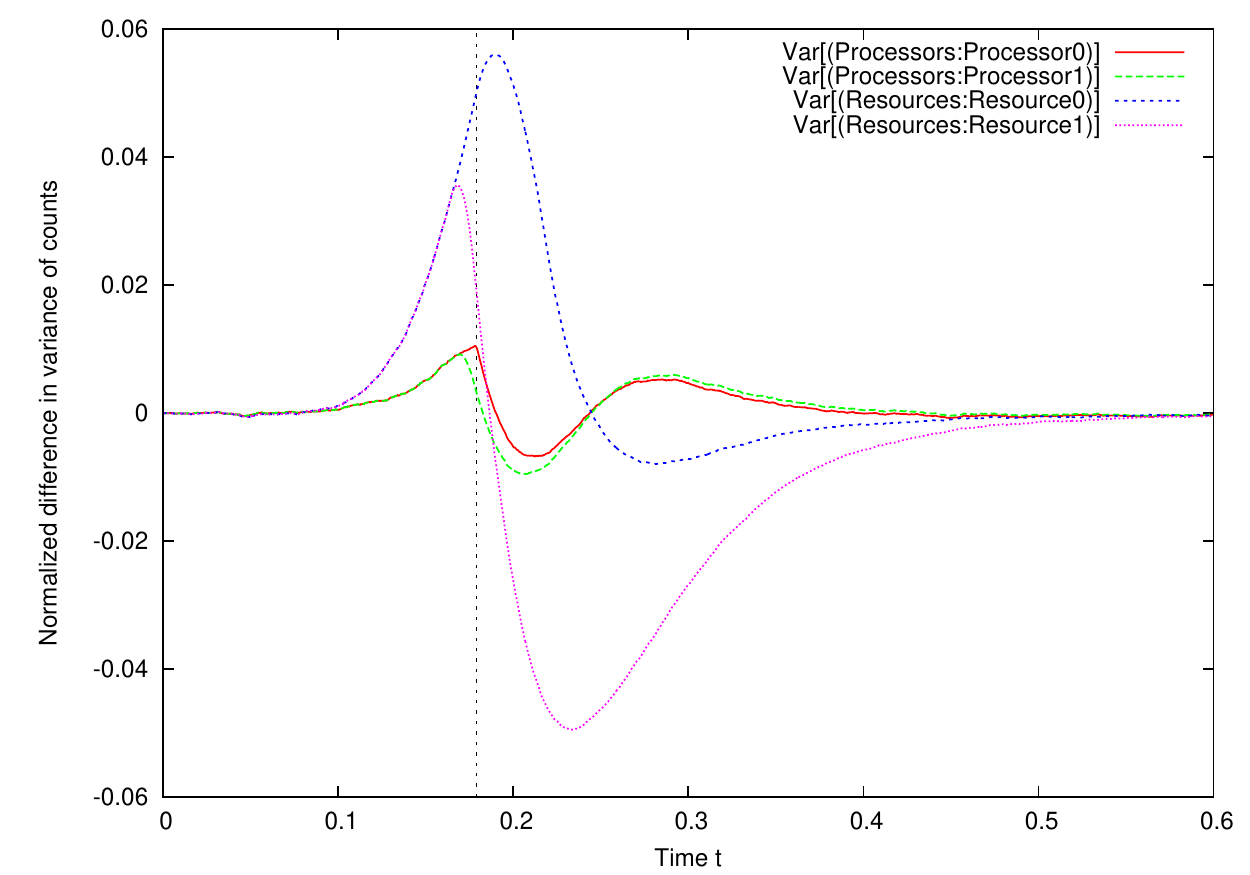}}
\subfloat[Normalised difference in variance for the scaled
version]{\includegraphics[scale=0.60]{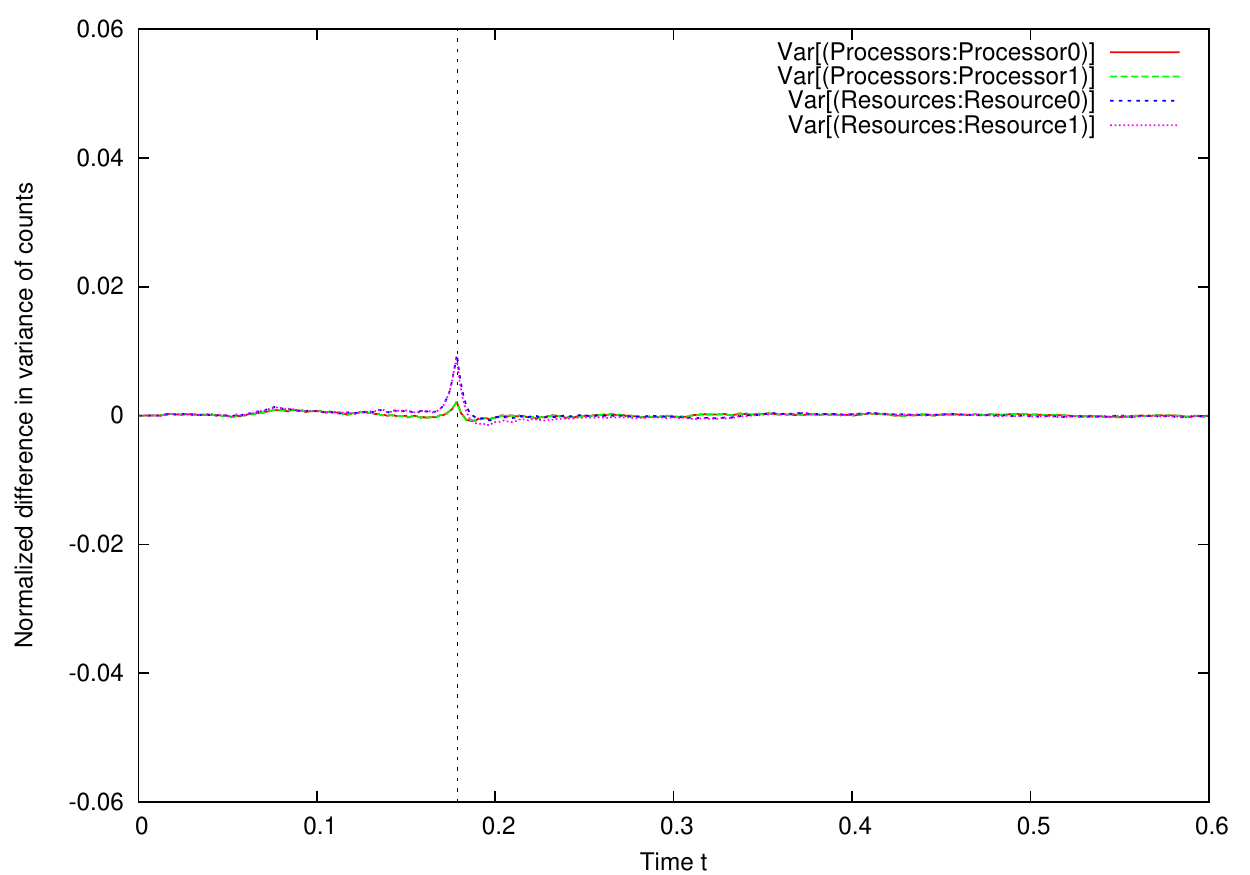}}
\end{center}
\caption{Difference in variance between the simulation and
its ODE approximation for the processor/resource model, produced by the
comparison analysis.}
\label{fig:procres_dif}
\end{figure}

Figures \ref{fig:procres_exp}, \ref{fig:procres_var} and
\ref{fig:procres_switch} that we used in the initial example in
\sectref{sec:example} were all produced by the respective \GPEPA\
commands enclosed in simulation and ODE analyses.  We could examine the
error of the approximation more closely by using the comparison
analyses. \figref{fig:procres_dif} compares the error in variance
approximation for the processors resource model with $m=50$ and $n=20$
with a version with populations scaled by $100$, \ie with $m=5000$ and
$2000$. In order to demonstrate the results from
\sectref{sec:theoretical}, the difference is scaled by the population
size $m+n$. We can see that under this scaling, the relative error
does decrease.

\clearpage
\section{\GPEPA\ syntax definition}

\def\syntax#1{\textcolor{blue}{\mathtt{#1}}}
\label{appendix:gpepa_syntax}
\textbf{Models}
\mbox{}
\begin{align*}
 \mathit{System} &:= \mathit{ParameterDefinition}^\ast ~
 \mathit{ComponentDefinition}^\ast ~
 \mathit{ModelDefinition} ~ \mathit{Analysis}^\ast\\
 &\\
 \mathit{ParameterDefinition} &:= \mathit{parameterId}~\syntax{=}~\mathit{realnumber}
 \syntax{;}\\
 &\\
 \mathit{ComponentDefinition} &:=
 \mathit{componentID}~\syntax{=}~\mathit{Component}\\
 \mathit{Component} &:= \mathit{Component}
\syntax{<}\mathit{ActionList}^?\syntax{>} \mathit{Component} \\
    &\phantom{:=}~\mid \mathit{Summation} \mid \mathit{componentId} \mid
    \syntax{(} \mathit{Component} \syntax{)}\\
 \mathit{Summation} &:= \mathit{Prefix} (\syntax{+} \mathit{Prefix})^\ast\\
\mathit{Prefix} &:=
\syntax{(}\mathit{actionId}\syntax{,}~\mathit{parameterId}~\syntax{).}(\syntax{(}\mathit{Summation}\syntax{)}\mid\syntax{stop}\mid\mathit{componentId}\mid)
 \syntax{;}\\
 &\\
 \mathit{ModelDefinition} &:=
 \syntax{(}\mathit{ModelDefinition}\syntax{<}\mathit{ActionList}^?\syntax{>}\mathit{ModelDefinition}\syntax{)}\\
 &\phantom{:= }~\mid\mathit{groupLabel}\syntax{\{}\mathit{ComponentsParallel}\syntax{\}}\\
 \mathit{ComponentsParallel} &:= \mathit{Component} (~\syntax{|}
 \mathit{Component})^\ast\\
 \mathit{Component} &:=
 \mathit{componentId}~(\syntax{[}\mathit{parameterId}\syntax{]})^?\\
 \mathit{ActionList} &:= \mathit{actionId}~(\syntax{,}~\mathit{actionId})^\ast
 \end{align*}
 \mbox{}
 \textbf{Analyses}
 \begin{align*}
 \mathit{Analysis} &:=
 \mathit{ODEs}~|~\mathit{Simulation}\mid\mathit{Comparison}\\
 \mathit{ODEs} &:= \syntax{odes(stopTime = }\mathit{realnumber}\syntax{,
 stepSize = }\mathit{realnumber}\syntax{, density = }\mathit{integer}\syntax{)}\\
        &\phantom{:=}~~~\syntax{\{}\mathit{Command}^\ast \syntax{\} }\\
 \mathit{Simulation} &:= \syntax{simulation(stopTime = }\mathit{realnumber}\syntax{,
 stepSize = }\mathit{realnumber}\syntax{, replications = }\mathit{integer}\syntax{)}\\
        &\phantom{:=}~~~\syntax{\{}\mathit{Command}^\ast \syntax{\} }\\
 \mathit{Comparison} &:=
 \syntax{comparsion(}\mathit{ODEs}\syntax{,}\mathit{Simulation}\syntax{)\{}\mathit{Command}^\ast\syntax{\}}
\end{align*}
\textbf{Commands}
\begin{align*}
 \mathit{Command} &:=
 \mathit{CommandNoFile}(\syntax{\texttt{->}"}\mathit{filename}\syntax{"})^?\syntax{;}\\
 \mathit{CommandNoFile} &:=
 \syntax{plot(}\mathit{MomentExpressions}\syntax{)}
 \mid\syntax{plotSwitchpoints(}\mathit{integer}\syntax{)}\\
 \mathit{MomentExpressions} &:= \mathit{MomentExpression}
 (\syntax{,}~\mathit{MomentExpression})^\ast\\
 \mathit{MomentExpression} &:=
 \mathit{MomentExpression}(\syntax{+}\mid\syntax{-}\mid
 \syntax{*}\mid\syntax{/}\mid\syntax{\textrm{\^{}}})\mathit{MomentExpression}\\
    &\phantom{:=}~\mid
    \syntax{E[}\mathit{Moment}(\syntax{+}\mathit{Moment})^\ast\syntax{]} \mid
    \syntax{Var[}\mathit{GCPair}(\syntax{+}\mathit{GCPair})^\ast\syntax{]} \\
    &\phantom{:=}~\mid (\syntax{Standardised})^?\syntax{Central[}\mathit{GCPair}\syntax{,}\mathit{integer}\syntax{]}\\
    &\phantom{:=}~\mid  \mathit{realnumber} \mid \mathit{parameterId} \mid
    \syntax{(} \mathit{MomentExpression} \syntax{)}\\
  \mathit{Moment} &:=
  (\mathit{GCPair}(\syntax{\textrm{\^{}}}\mathit{integer})^?)^{+} \\
  \mathit{GCPair} &:= \mathit{groupLabel}\syntax{:}\mathit{Component} 
\end{align*}

\end{document}